\documentclass{sig-alternate-05-2015}
\usepackage[english]{babel}

\usepackage{amsmath}
\usepackage{amsfonts}
\usepackage{amssymb}
\usepackage{bbm}

\usepackage{url}
\usepackage{hyperref}

\usepackage{booktabs}
\usepackage{rotating}
\usepackage{multirow}
\usepackage{array}

\usepackage{verbatim}
\usepackage{graphicx}
\usepackage{subfig}
\usepackage{epstopdf}
\DeclareGraphicsExtensions{.pdf,.eps,.png,.jpg,.tif,.tiff,.ps}
\graphicspath{{.}{img/}}

\usepackage[svgnames]{xcolor}

\usepackage{pdfpages}
\usepackage{xspace}
\newcommand{\etal}{{\em et al.}\xspace}
\definecolor{navy}{rgb}{0.1, 0.1, 0.8}
\definecolor{gray}{rgb}{0.4, 0.4, 0.4}
\definecolor{myblue}{rgb}{.8, .8, 1}
\definecolor{olive}{rgb}{0.1, 0.5, 0.1}

\newcommand{\eat}[1]{}

\newcommand{\rvx}[1]{{#1}}

\usepackage{soul} 

\newcommand{\seismic}{{\sc Seismic}\xspace}
\newcommand{\news}{{\sc News}\xspace}
\newcommand{\seismicdata}{{\sc Tweet-1Mo}\xspace}
\newcommand{\hawkes}{{\sc Hawkes}\xspace}
\newcommand{\hawkesC}{{\sc HawkesC}\xspace}
\newcommand{\hybrid}{{\sc Hybrid}\xspace}
\newcommand{\featuredriven}{{\sc Feature-Driven}\xspace}

\newcommand{\eqmoveup}{\vspace{-0.0in}}                 
\newcommand{\captionmoveup}{\eqmoveup\vspace{-0.14in}}   

\newcommand{\squishlisttwo}{
 \begin{list}{$\bullet$}
  { \setlength{\itemsep}{0pt}
    \setlength{\parsep}{0pt}
    \setlength{\topsep}{0pt}
    \setlength{\partopsep}{0pt}
    \setlength{\leftmargin}{1.5em}
    \setlength{\labelwidth}{1.5em}
    \setlength{\labelsep}{0.5em} } }

\newcommand{\squishend}{
  \end{list}  }
\newcommand{\squeezeup}{\vspace{-2.0mm}}

\hypersetup
{
	pdftitle={Feature Driven and Point Process Approaches for Popularity Prediction},%
	pdfauthor={Swapnil Mishra, Marian-Andrei Rizoiu and Lexing Xie},%
	pdfkeywords={social media; self-exciting point process; information diffusion; cascade prediction.}%
}

\begin{document}

\CopyrightYear{2016}
\setcopyright{acmcopyright}
\conferenceinfo{CIKM'16 ,}{October 24-28, 2016, Indianapolis, IN, USA}
\isbn{978-1-4503-4073-1/16/10}
\acmPrice{\$15.00}
\doi{http://dx.doi.org/10.1145/2983323.2983812}

\title{Feature Driven and Point Process Approaches for Popularity Prediction}


%
%
%
%
%

\numberofauthors{3}
\author{
    \alignauthor Swapnil Mishra\\
    \alignauthor Marian-Andrei Rizoiu\\
    \alignauthor Lexing Xie\\
    \and
    The Australian National University, Data 61, CSIRO, Australia\\
    \and
	\{swapnil.mishra,marian-andrei.rizoiu,lexing.xie\}@anu.edu.au
}

\eat{
\numberofauthors{3} 
%
\author{
%
%
\alignauthor
Swapnil Mishra\\
       \affaddr{ANU \& Data61\\ Canberra, Australia}\\
       \email{swapnil.mishra@anu.edu.au}
\alignauthor
Marian-Andrei Rizoiu\\
       \affaddr{ANU \& Data61, Canberra}\\
       \email{marian-andrei.rizoiu\\@anu.edu.au}
\alignauthor Lexing Xie\\
       \affaddr{ANU \& Data61\\ Canberra, Australia}\\
       \email{lexing.xie@anu.edu.au}
}
}

\maketitle
\begin{abstract}

Predicting popularity, or the total volume of information outbreaks, 
is an important subproblem for understanding collective behavior in networks. 
Each of the two main types of recent approaches to the problem, feature-driven and generative models, 
have desired qualities and clear limitations. 
This paper bridges the gap between these solutions 
with a new hybrid approach and a new performance benchmark.
We model each social cascade with a marked Hawkes self-exciting point process,  
and estimate the content virality, memory decay, and user influence. 
We then learn a predictive layer for popularity prediction 
using a collection of cascade history. 
To our surprise, Hawkes process with a predictive overlay outperform
recent feature-driven and generative approaches 
on existing tweet data~\cite{Zhao2015} and a new public benchmark on news tweets.
We also found that a basic set of user features and event time summary statistics
performs competitively in both classification and regression tasks, 
and that adding point process information 
to the feature set further improves predictions.
From these observations, we argue that future work on popularity prediction 
should compare across feature-driven and generative modeling approaches in 
both classification and regression tasks. 

\end{abstract}

%
%

%

%
%

%
%


\vspace{3mm}\noindent{\bf Keywords} social media; self-exciting
point process; information diffusion; cascade prediction.


\section{Introduction}

Popularity prediction, especially in the context of online media, has seen increased attention, as
successful solutions would help both content consumers to cope with information overload, and content producers to disseminate information within limited resources. 
Accurate models will help producers to identify the trends, discover useful content and decimate it faster to improve the content delivery of the network. 
Furthermore, this would provide with more insights into understanding collective behaviour by compiling seemingly incoherent individual responses to a single piece of information. 

When presented with the task of predicting popularity, a clear gap appears between the two main classes of approaches: the feature-driven approaches and the generative approaches.
\emph{Feature-driven approaches}~\cite{Asur2010,CHE14,Martin2016,Pinto2013} summarize network, user, and event history information into an extensive set of features, and they use machine learning approaches to predict future popularity.
\emph{Generative methods}~\cite{Ding2015,Shen2014,yu2015micro,Zhao2015} leverage fine-grained timing information in the event series, however they make stronger assumptions about the diffusion process.
A thorough understanding of strengths and weaknesses of the two classes is currently missing, especially since the generative methods are usually designed for explaining the mechanisms that generate attention, and not optimized for prediction. 
Another gap arises between the different problem settings.
Prior work has address problems like
predicting the volume of attention~\cite{Shen2014,Zhao2015}, 
predicting if a cascade will double in size~\cite{CHE14}, 
whether an item will have 10 million views~\cite{shamma2011viral}, or be among the top 5\% most popular~\cite{yu2014twitter}. 
However, the understanding of how the different approaches and the different features generalize over the various problem settings is incomplete.  
In addition, a practical bottleneck is the availability of certain types of these features, such as network data and corresponding features -- for example the number of users exposed in the diffusion up to a time $t$ or potential number of reachable users. 
A third gap in the current available work is the consensus and understanding about which features are informative for feature-driven models and the lack of benchmark datasets on which new methods can be devised.
In this paper, we address the above challenges in the context of 
predicting the final size of retweet cascades.
We build two predictive approaches, one generative and one feature-driven, and we show how to combine them into a hybrid predictor to further increase performances.
The generative method is a two-layered approach, built on the intuitive \emph{Hawkes self-exciting process}~\cite{HAWKES1971} model.
Self-exciting point processes are a class of stochastic processes, in which the occurrence of past events makes the occurrence of future events more probable.
Three key factors in information diffusion are built into the proposed model: the social influence of users, the length of ``social memory'' and the inherent tweet quality.
We use a predictive layer on top of the generative model to make final predictions. 
This helps us to take into account other cascades and mitigate limitations of model assumptions and parameter estimation.
Our second proposed approach, the feature-based method, uses
only features which can be computed on data containing solely the message content and basic user profiles, and which were highlighted as most predictive by recent studies: user related features~\cite{Martin2016} and temporal activity features~\cite{CHE14}.
This results in a  competitive feature-based predictor, which consistently outperforms the current state-of-the-art popularity prediction model~\cite{Zhao2015} -- reducing the mean absolute relative error by more than 200\% when compared to the latter, after observing the retweet series for 10 minutes.
We show that the same set of features can be employed in both regression and classification tasks.
We further experiment with a hybrid predictor, which uses both data features and measures issued from the generative model,
and we show additional performance improvement.
\begin{figure}[tbp]
	\includegraphics[width=0.48\textwidth]{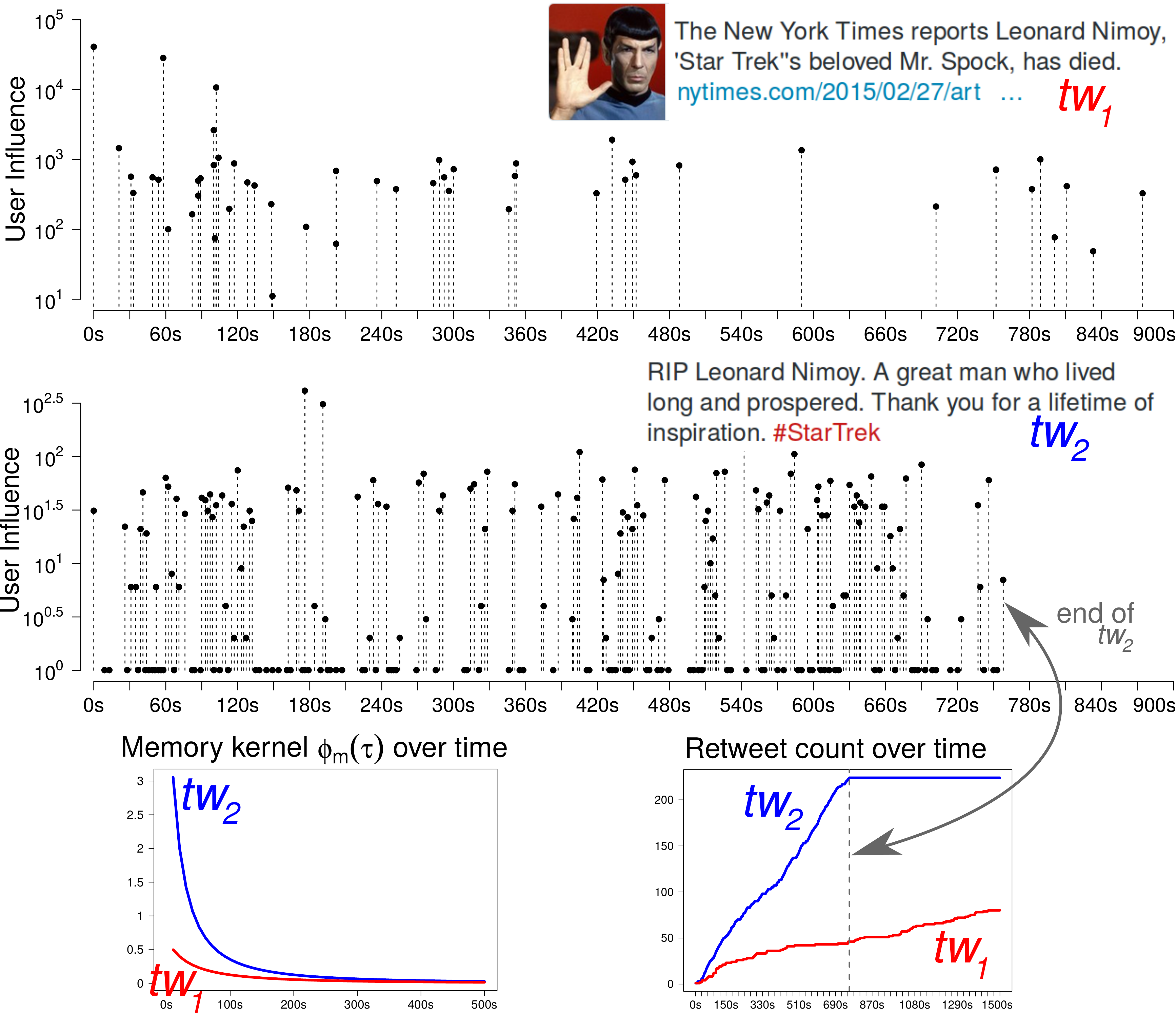}
    \caption{
	Two retweet cascades announcing the death of ``Mr. Spock''.
	The underlying diffusion processes of $tw_1$ and $tw_2$ are very different: 
	$tw_1$ lasts for 2.5 hours. It attracts the attention of highly influential users, especially music related channels, however the generative kernel of $tw_2$ amplifies more than that of $tw_1$.
	$tw_2$ diffuses	faster and ends in just 12 minutes, but the final sizes for the two cascades are similar (224 retweets for $tw_1$ and 219 for $tw_2$).
	We model and interpret these cascades using our generative model in Sec.~\ref{subsec:model-interpretation}.
	x-axis: time in seconds; y-axis: number of followers each tweet can reach; $\phi_m(\tau)$: memory kernel of Eq.~\eqref{eq:phi-separable} plotted with $m=1000$. 
    }
    \label{fig:example-cascade}\captionmoveup
\squeezeup
\end{figure}
%
We conduct experiments on a large dataset, containing 49.7 million tweets containing links to top 10 news sites in English, gathered using the public API over a period of 4 months, in 2015. 
Fig.~\ref{fig:example-cascade} illustrates two of the retweet cascades in this dataset, relating the passing of the actor for ``Mr. Spock''.
Even though the two cascades achieve similar popularities, our generative model is capable of differentiating between their very different diffusion dynamics.
For example, the memory kernel $\phi_m(\tau)$ of $tw_2$ -- i.e. the contribution of each event to the spawning of future events -- amplifies more (shown by a larger area under the $\phi_m(\tau)$), but is forgotten faster (the entire diffusion ends in just 12 minutes).


There are three main contributions of this paper:
\squishlisttwo
	\item \textbf{Bridging the popularity prediction problem space:}
	we address three types of gaps: problem settings and task (e.g. regression vs. classification), type of approach (feature-based vs. generative) and the generalization of the features across feature-driven approaches.
	
	\item \textbf{Comparative understanding of feature-driven and generative models:} 
	we propose a generative method, based on Hawkes self-exciting processes, which features the advantages of generative methods -- e.g. interpretable results -- and which is adapted for predicting the final cascade popularity by using an additional predictive layer; 
	we compare approaches to popularity prediction from the two main classes, i.e. feature-based and generative, and we show that combining them into a hybrid approach increases performances.

	\item \textbf{Dataset:}  
	we identify a set of features that generalize over different settings and can be constructed with only message content and basic user features; 
	we curate a large, domain-specific dataset of retweet diffusion cascades, which can be used by both feature-driven and generative models for prediction. 
	Dataset along with code is publicly available for download.\footnote{Download data and code here: \url{https://git.io/v6rIN}.} 
	This new dataset allows comprehensive performance benchmarks on features about users and temporal activity.
\squishend

The rest of this paper is structured as follows.
Sec.~\ref{sec:related-work} presents related work.
Sec.~\ref{sec:model} details the proposed generative model and fitting methods, while Sec.~\ref{sec:prediction} presents how the popularity prediction is performed, based on the fitted model parameters.
The dataset and feature construction for the feature-driven approach are detailed in Sec.~\ref{sec:data} and the experiments are outlined in Sec.~\ref{sec:xp}.
We conclude in Sec.~\ref{sec:conclusion}.

\section{Related Work}
\label{sec:related-work}


We structure the discussion of related work onto the
two broad previously mentioned categories: \emph{feature-driven approaches} and \emph{generative approaches}.
For generative models, being somewhat recent,  we discuss their application in popularity prediction as well as their use in other social media settings. 

\textbf{Predicting popularity with feature-driven models.}
Data-driven approaches~\cite{Davletov2014,Kamath2013} treat popularity as a non-decomposable process and take a bottom-up approach.
They rely on machine learning algorithms to make the connection between item popularity and an extensive set of features.
Most often the main contributions of such approaches are collecting large-scale datasets and constructing informative features.
Bakshy \etal~\cite{BAK11} and later Martin \etal~\cite{Martin2016} predict tweet popularity, defined as the total size of its retweet cascade.
Both studies construct features relating to the user having posted the tweet and posted content.
User features, particularly past user success -- the ability of a user to start large cascades -- are shown more informative, however prediction performances are limited.
Martin \etal make the hypothesis of an inherent popularity randomness, which sets limits to its predictability.
The situation is different when part of the diffusion cascade is observable, as early activity has been shown predictive for total popularity~\cite{Pinto2013,szabo2010}. 
When predicting whether a Facebook cascade will double in size, Cheng \etal~\cite{CHE14} found that temporal features -- like the resharing rate or the acquired volume of views -- are the most informative features.
Prediction performances are further improved when adding features which account for specific contexts of the information diffusion.
For example, Li \etal~\cite{Li2013a} add propagation network measurements, Chang \etal~\cite{Chang2014} add sophisticated features relating to audience behavior when predicting the popularity of serials and Vallet \etal~\cite{Vallet2015} construct features relating to two social networks: Youtube and Twitter.
%
Many proposed features cannot be used outside of the proposed context, 
being either data-source specific 
or relying on information that is not publicly accessible.
The feature-driven methods that we propose use only the best-performing features shown 
in previous studies and which can be extracted from public data.

%

\textbf{Generative models in social media.}
Generative models start by modeling a set of phenomena into a parametric model and fitting the parameters of the model to observed data.
Such approaches have been successfully used in neuro-biology to model neuronal spiking activity~\cite{Kim2011}, in economics to model the evolution of stock prices~\cite{Filimonov2013}, in physics to model earthquake aftershocks occurrence \cite{SOR99,Helmstetter2002,Sornette2003} and in criminology to model crime patterns~\cite{Mohler2011}.
In social media such approaches have been used extensively~\cite{JOH00,pinto2015trend,raghavan2014modeling}, mainly due to their capability to model phenomena like user interactions and diffusion susceptibility.
Hidden network properties, such as connexions between individuals~\cite{li2013dyadic,li2014learning} or the evolution of the diffusion network topology~\cite{farajtabar2015coevolve}, have been uncovered using generative methods, such as self-exciting point processes. 
The main difficulty with such approaches is scalability, since each social phenomenon must be accounted manually in the model.

\textbf{Modeling and predicting popularity using generative models.}
Popularity modeling~\cite{Crane2008,Ding2015,yu2015micro} and prediction~\cite{Shen2014,Zhao2015} has been a particularly fertile field for point-process based generative models.
In their seminal work, Crane and Sornette~\cite{Crane2008} showed how a Hawkes point-process can account for popularity bursts and decays.
Afterwards, more sophisticated models have been proposed to model and simulate popularity in microblogs~\cite{yu2015micro} and videos~\cite{Ding2015}.
These approaches successfully account for the social phenomena which modulate online diffusion: the ``rich-get-richer'' phenomenon and social contagion.
Certain models can output an estimate for the total size of a retweet cascade.
Shen \etal~\cite{Shen2014} employ reinforced Poisson processes, modeling three phenomena: fitness of an item, a temporal relaxation function and a reinforcement mechanism.
Zhao \etal~\cite{Zhao2015} propose SEISMIC, which employs a double stochastic process, one accounting for infectiousness and the other one for the arrival time of events.
It is the current state of the art in popularity prediction and, throughout this paper, we compare against it.
Our proposed generative model uses the same information as SEISMIC -- tweet timestamps and the number of followers of the user posting the tweet --, but uses a simpler, more intuitive modeling of the diffusion cascade.

Generative models are typically designed to explain popularity, not predict it.
Most of the methods presented earlier introduce multiple regularization and correction factors 
in order to make meaningful predictions.  
This is sub-optimal both for prediction and for interpretation. 
By contrast, our Hawkes predictive model separates modeling from prediction
accounting directly for the most important three factors at individual cascade level: user influence, decaying social attention and content quality, 
followed by a predictive layer tuned from historical data. 
\section{Retweeting as a Point Process}
\label{sec:model}

We first introduce the Hawkes self-exciting point process, a stochastic event model for describing retweeting cascades.
We then describe a procedure for maximum-likelihood estimation of model parameters from an observed cascade. 
We present observations of the Hawkes model on individual and collections cascades, illustrating the interplay of individual memory kernel and user influence in diffusion dynamics.

\subsection{A self-exciting point processes}

In self-exciting processes the occurrences of past events make future events more probable. 
In particular, we model each retweet as a stochastic event, and model three key intuitions in a social network: 
\emph{magnitude of influence}, tweets by users with many followers tend to get retweeted more; 
\emph{memory over time}, that most retweeting happens when the content is {\em fresh}~\cite{Wu2007};
and \emph{content quality}.

The marked Hawkes process~\cite{HAWKES1971} is a well-known self-exciting process that captures these properties. 
Its event rate is written as: 
\begin{equation} \label{eq:hawkes}
	\lambda(t) = \sum_{t_i < t} \phi_{m_i}(t - t_i)\enspace,
\end{equation}
where $\lambda(t)$ is the arrival rate of new events. 
It depends 
on each previously occurred event $(m_i, t_i)$ of magnitude $m_i$ at time $t_i$
through the triggering kernel $\phi_m(\tau)$. 
For the problem of estimating the size of a retweet cascade, there are no exogenously generated events (as in a general Hawkes process~\cite{Hawkes1974}) apart from the initial tweet, which is accounted for with magnitude $m_0$ at time $t_0=0$.
The effects of $m_i$ and $t_i$ are separable in our triggering kernel:
\begin{equation}
	\phi_m(\tau) = \kappa m^{\beta} (\tau + c)^{-(1+\theta)} \enspace.
	\label{eq:phi-separable}
\end{equation}
$\kappa$ describes the {\em virality} -- or quality -- of the tweet \rvx{content} and it scales the subsequent retweet rate; 
$\beta$ introduces a warping effect for the user influence and it is related to the observed long-tail distribution of user influence in social networks;
and $1+\theta$ ($\theta>0$) is the power-law exponent, describing how fast an event is {\em forgotten}, parameter $c>0$ is a temporal shift term to keep $\phi_m (\tau)$ bounded when $\tau \simeq 0$ ; 
Here $\kappa m^{\beta}$ accounts for the magnitude of influence, and the power-law kernel $(\tau + c)^{-(1+\theta)}$ models the memory over time.
We approximate user influence $m$ using the number of followers
obtained from Twitter API.
Table~\ref{tab:parameters} summarizes the parameters of the model, along with their interpretations.

Throughout literature, three families of functions have been mainly used to model temporal decay with Hawkes point-processes~\cite{Gomez-Rodriguez2011}: 
power-law functions $\phi^{p}(\tau) = (\tau + c)^{-(1+\theta)}$, used in geophysics~\cite{Helmstetter2002} and social networks~\cite{Crane2008} ; 
exponential functions $\phi^{e}(\tau) = e^{-\theta \tau}$, used in 
financial data~\cite{Filimonov2013};
Reyleigh functions  $\phi^r(\tau) = e^{-\frac{1}{2}\theta\tau^2}$, used in epidemiology~\cite{Wallinga2004}.
We experimented with all three kernels with the setting in~Sec.~\ref{sec:xp}, and found that
the power-law kernel consistently outperform the other two. 
Consequently, we only discuss detailed results from the power law kernel in the rest of this paper.


\begin{table}[!tb]
\caption{\rvx{Summary of notations.
Top: Quantities for estimating popularity using point processes.
Bottom: Parameters of the marked Hawkes process.}}
\small
\centering
\begin{tabular}{cp{6.5cm}}
\toprule
Notation & Interpretation \\ 
\midrule
$m_i$ & event magnitude. Social influence of the user emitting the tweet. \\ 
$t_i$ & event time. Tweet timestamp. \\ 
$\lambda(t)$ & conditional intensity -- or event rate -- of a non-homogeneous Poisson Process. \\ 
$\phi_m(\tau)$ & triggering kernel. Contribution of event the $(m, t)$ to the total event rate, calculated at time $t + \tau$. \\ 
$n^{\ast}$ &  branching factor, mean number of children events spawned by a parent event. For $n^{\ast} < 1 $ the cascade dies out (subcritical regime), for $n^{\ast} > 1$ the cascade explodes exponentially (supercritical regime). \\ 
$T$ & time extent of the observed series of events in the beginning of a retweet cascade. $T = max(t_i)$. \\
$n$ & number of events in the observed series of events in the beginning of a retweet cascade. $n = max(i)$. \\
$N_\infty$ & expected total number of events in a given retweet cascade, derived from the generative model. \\
$N_{real}$ & the real number of events in a given retweet cascade, or total popularity.\\
$\omega$ & predictive factor learned using a Random Forest regressor\eat{ in the predictive layer}. \\
$\hat N_\infty$ & total popularity prediction made using the output of the predictive layer.\\
\bottomrule
 &  \\ 
\toprule
Parameter & Interpretation \\ 
\midrule
$\theta$ & the power-law exponent, describing how fast an event is {\em forgotten}. $\theta > 0$. \\ 
$\kappa$ & tweet quality. High quality tweets are more likely to generate more retweets. $\kappa > 0$. \\ 
$c$ & temporal shift cutoff term so that $\phi_m (\tau)$ stays bounded when $\tau \simeq 0$. $c > 0$. \\ 
$\beta$ & user influence power-law warping effect in $b(m)$. $\beta > 0$. $0 < \beta < \alpha - 1$ for having a defined branching factor $n^*$.  \\ 
$\alpha$ & exponent of the user influence power-law distribution. Fitted to the \texttt{\#followers} distribution: $\alpha = 2.016$. \\ 
\bottomrule
\end{tabular}
\label{tab:parameters}\captionmoveup
\end{table}

\subsection{The branching factor}
\label{subsec:branching-factor}

One of the key 
quantities that describe a Hawkes process is the branching factor $n^{\ast}$, defined as the expected number of children events directly spawned by an event. 
Note that the total number of expected events from one initial event is the sum of a geometric series $\sum_{k=1}^{\infty} (n^{\ast})^k$. 
For $n^{\ast} < 1$, the process in a \emph{subcritical regime}: the total number of retweets is bounded, the event rate $\lambda(t)$ decays to zero over time and the retweet cascade dies out.
When $n^{\ast} > 1$, the process is in a so-called \emph{supercritical regime} with the total number of retweets being unbounded.
%
We compute the branching factor by integrating over time and event magnitudes.
\begin{equation} \label{eq:nstar-general}
	\eqmoveup
	n^{\ast} = \int_{1}^\infty \int_{0}^\infty P(m) \phi_{m}(\tau) d\tau dm \enspace.
\end{equation}
\eat{where $P(m)$ is the distribution over user influence from which we assume that $m_i$ is sampled. In the case of social networks, user influence has been shown to follow a power-law distribution~}
Here we assume $m_i$ are {\em i.i.d.} samples from a power law distribution of social influence~\cite{kwak2010twitter}: $P(m) = (\alpha - 1) \: m^{-\alpha}$.
Here $\alpha$ is an exponent which controls the heavy tail of the distribution.  
We extract the number of followers $m$ for a large sample of users from our dataset described in Sec.~\ref{subsec:dataset} and fit it to a power-law distribution following the method detailed in~\cite{Clauset2014}. 
We obtain $\alpha = 2.016$ and we use it throughout the experiments.
Substituting Eq.~\eqref{eq:phi-separable} and $P(m)$ into~\eqref{eq:nstar-general}, we obtain the closed-form expression of the branching factor:
\begin{equation}
	\eqmoveup
	n^{\ast} = \kappa \frac{\alpha - 1}{\alpha - \beta - 1} \frac{1}{\theta c^\theta}, \text{  for } \beta < \alpha - 1 \text{ and } \theta > 0 \enspace.
\end{equation}

Not only the branching factor $n^{\ast}$ is an intuitive descriptor of the {\em virality} of a cascade, it is also used for predicting the total popularity in Sec~\ref{sec:prediction} and as a constraint in model estimation in Sec~\ref{ssec:loglikelihood}.

\vfill\eject
\subsection{\rvx{Estimating the Hawkes process}}
\label{ssec:loglikelihood}

{\bf The log-likelihood function.}
Our marked Hawkes process
is completely defined by the four parameters $\{ \kappa, \beta, c, \theta\}$.
We now describe a maximum-likelihood estimation procedure of the model from a set of observed events, consisting of $\{(m_i,t_i), i=0,\ldots,n\}$ for time $t<T$.
%
The log-likelihood function of a point-process described by Eq.~\eqref{eq:hawkes} and~\eqref{eq:phi-separable} is written as (\cite{Daley2008}, Ch.~7.2):
\begin{align}
{\cal L}(\kappa, \beta, c, \theta) =& \log P(\{(m_i, t_i), i=1,\ldots,n\}) \nonumber \\
= &\sum_{i=1}^n \log\left(\lambda\left(t_i\right)\right) - \int_0^T \lambda(\tau)\mathrm{d}\tau \nonumber \\
^{cf.~\eqref{eq:hawkes} \& \eqref{eq:phi-separable}} = &\sum_{i=2}^n\log\kappa + \sum_{i=2}^n \log\left( \sum_{t_j<t_i}\dfrac{\left(m_j\right)^\beta}{\left(t_i - t_j + c\right)^{1+\theta}}\right) \nonumber \\ 
&- \kappa \sum_{i=1}^n \left({m_i}\right)^\beta \left[\dfrac{{1}}{\theta c^{\theta}} - \dfrac{\left(T+c-t_i\right)^{-\theta}}{\theta} \right] \enspace. \label{eq:ll}
\end{align}
The first two terms in Eq.~\ref{eq:ll} are from the likelihood computed using the event rate $\lambda(t)$, the last term is a normalizing factor from integrating the event rate over the observation window $[0,T]$.

{\bf Finding the maximum-likelihood solution.} 
Eq.~\ref{eq:ll} is a non-linear objective that we optimize to find the set of parameters that maximizes data likelihood. 
Directly maximizing this objective is more efficient than expectation-maximization approaches~\cite{Ding2015} that try to estimate the branching structure (of which event triggers which) and that have a quadratic number of latent variables. 
There are a few natural constraints for each of model parameter, namely: $\theta>0$, $\kappa>0$, $c>0$, and $0<\beta<\alpha-1$ for the branching factor to be meaningful (and positive).
Furthermore, while the supercritical regimes $n^{\ast} > 1$ are mathematically valid, they will not lead to a finite cascade size prediction. 
We note that in reality no retweet cascade can grow indefinitely, due to finite network sizes. 
We further incorporate $n^{\ast} < 1$ as a non-linear constraint for the maximum likelihood estimation.
We use the Ipopt~\cite{Wachter2006}, the large-scale interior point solver that handles both non-linear objectives and non-linear constraints, with pre-programmed gradient functions to find the solution. 
Details of the gradient computation and optimization can be found in an online supplement~\cite{supplemental}.

\subsection{Interpreting the generative model}
\label{subsec:model-interpretation}

\begin{figure}[tbp]
    \centering
    \subfloat[] {
		\includegraphics[height=0.115\textheight]{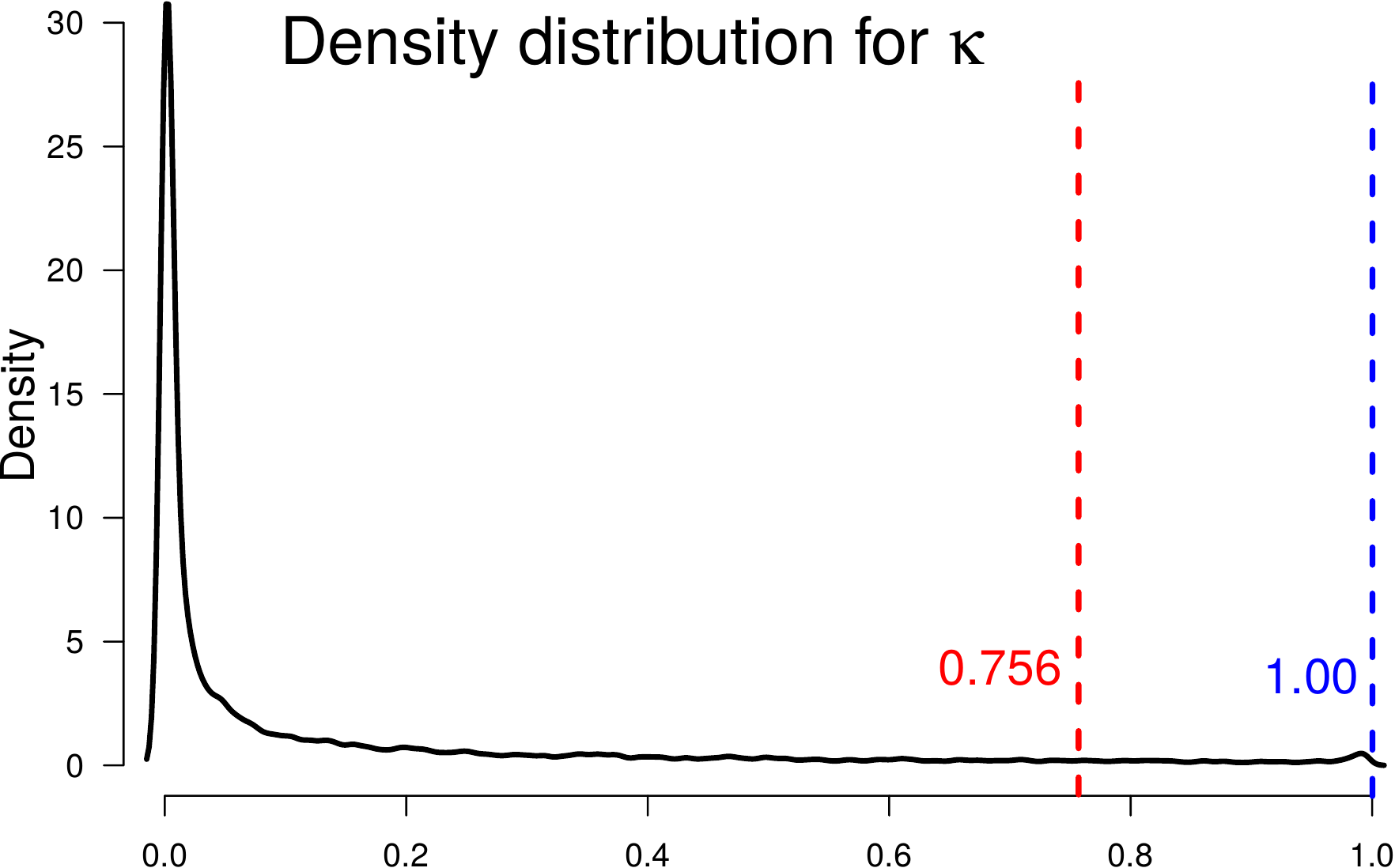}
	} 
	\subfloat[] {
		\includegraphics[height=0.115\textheight]{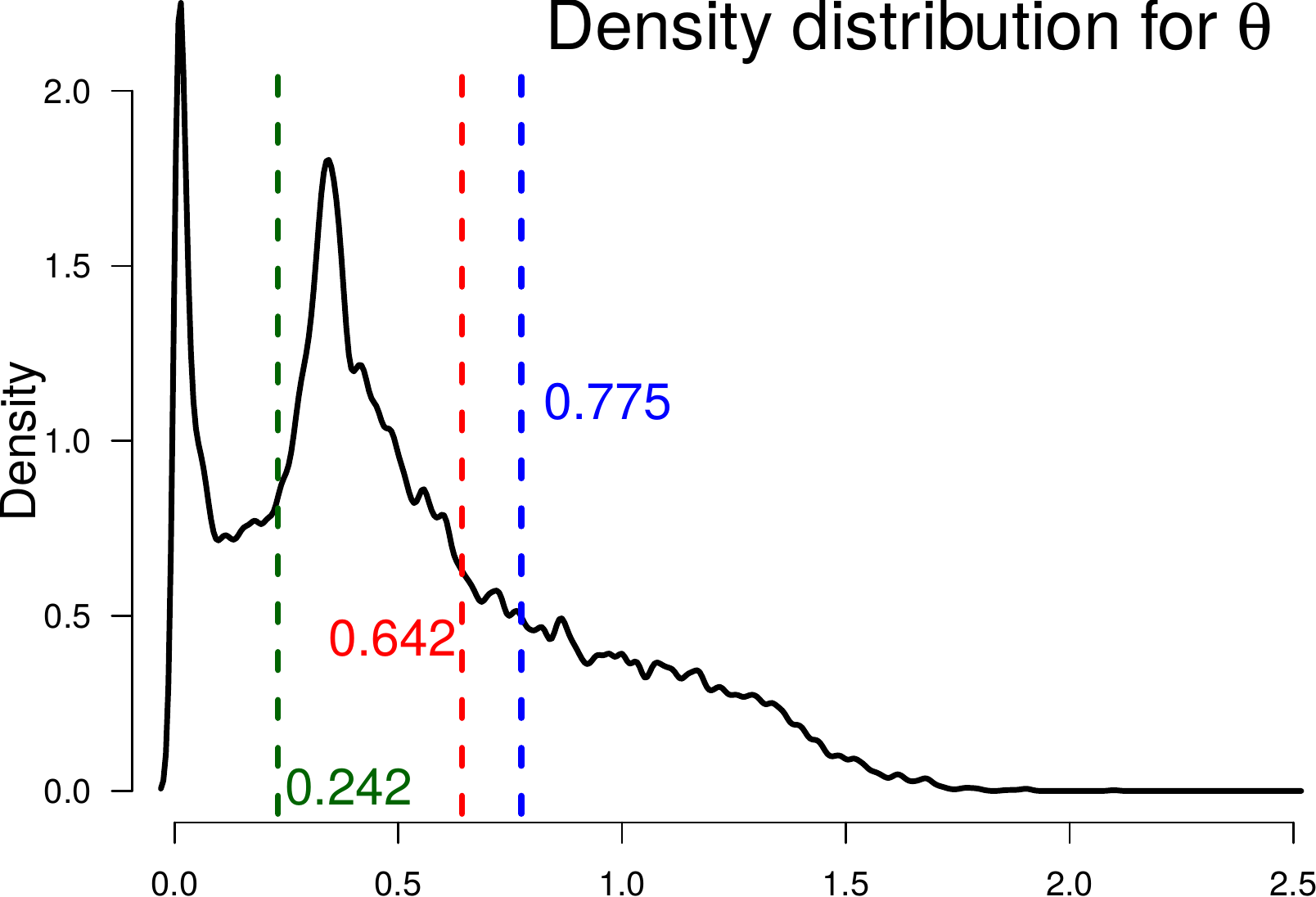}
	}
    \caption{
	Density distribution of model parameters $\kappa$ (a) and $\theta$ (b), on a sample of 17,146 cascades from the \news dataset.
	The dashed vertical lines show fitted parameter values for the two retweet cascades illustrated in Fig.~\ref{fig:example-cascade}: $tw_1$ in \textcolor{red}{red} and $tw_2$ in \textcolor{blue}{blue}.
	We note that \seismic~\cite{Zhao2015} uses a fixed value of $\theta = 0.242$ for all cascades (denoted in green).
	The distributions for $c, \beta$ and $n^*$ are shown in the supplement~\cite{supplemental}.
    }
    \label{fig:model-interpretation}
    \captionmoveup

\end{figure}

\textbf{Individual cascades: fast vs slow.}
We first examine the model parameters for the two cascades illustrated in Fig.~\ref{fig:example-cascade}. 
These two cascades are about the same event and have similar observed popularities (224 vs. 219) through very different diffusion speeds at a complex interplay between power-law memory $(\tau+c)^{-(1+\theta)}$ and user influences.
For $tw_1$, the maximum-likelihood estimate of parameters are $\{\kappa = 0.75, \beta = 0.27, c = 58.46, \theta =  0.64 \}$ with a corresponding $n^\ast = 0.12$;
for $tw_2$, $\{\kappa = 1, \beta = 0.42, c = 27.77, \theta =  0.77 \}$, and $n^\ast = 0.16$ (also shown in Fig.~\ref{fig:model-interpretation}).
The two diffusions unfold in different ways:
$tw_1$ has lower value of $\kappa$ and higher waiting time $c$, hence its memory kernel $\phi(\tau)$ has lower values than that of $tw_2$ and a slower decay.
However, $tw_1$ attracts the attention of some very well-followed music accounts.
The original poster (\texttt{@screencrushnews}) has 12,122 followers, and among the retweeters \texttt{@TasteOfCountry} (country music) has 193,081 followers, \texttt{@Loudwire} (rock) had 110,824 followers, \texttt{@UltClassicRock} (classic rock) has 99,074 followers and \texttt{@PopCrush} (pop music) has 114,050 followers. 
The resulting cascade reached 1/4 its size after half an hour, and the final tweet was sent after 4 days.
In contrast, $tw_2$'s memory kernel has higher values, but faster decay. 
The most influential user in $tw_2$ has 412 followers, and the entire diffusion lasted for only $12.5$ minutes.


\textbf{Interpreting a collection of cascades.}
We construct a large subset of 17,146 cascades of the \news dataset, described in Sec.~\ref{subsec:dataset}.
For each cascade we fit the parameters of its generative model $\{\kappa, \beta, c, \theta \}$.
Fig.~\ref{fig:model-interpretation} shows the density distribution of parameters $\theta$ and $\kappa$.
As one would expect, most cascades have a low value of $\kappa$, and reflect the long-tailed distribution of content quality. 
$n^\ast$ -- descriptor of the virality of a cascade -- is distributed very similarly to $\kappa$, with most cascades showing low $n^\ast$ values (shown in the supplement~\cite{supplemental}).
%
Parameter $\theta$ -- which controls for the speed of the decay of the memory kernel $\phi_{m}(\tau)$ -- has a more surprising distribution.
Recent work~\cite{Zhao2015} on point processes with power-law kernels argued that a unique value of $\theta = 0.242$ is sufficient to explain the general behavior cascades.
We find $\theta$ to vary in the interval $(0, 2.8]$ and showing two maxima at $0.01$ and $0.34$.
This observation indicates that learning individual memory exponents $\theta$
allows us to better describe the diffusion dynamics of each cascade.
The distributions of the other parameters are shown in the supplement: 
$c$ follows a long-tail distribution; 
$\beta$ shows a maximum of density around $0.08$ and a smaller peak around $\beta \sim \alpha - 1$ resulting from the optimizer terminating at the boundary of the non-linear constraint $n^\ast = 1$.
\section{Predicting Popularity}
\label{sec:prediction}

In this section, we 
first 
derive the number of expected future events in a cascade, 
we then describe a predictive layer tuned for estimating the total size, 
using information from other cascades in history. 

\subsection{The expected number of future events}
\label{final-retweets-estimation}

\begin{figure}[tbp]
\centering
\includegraphics[width=0.49\textwidth]{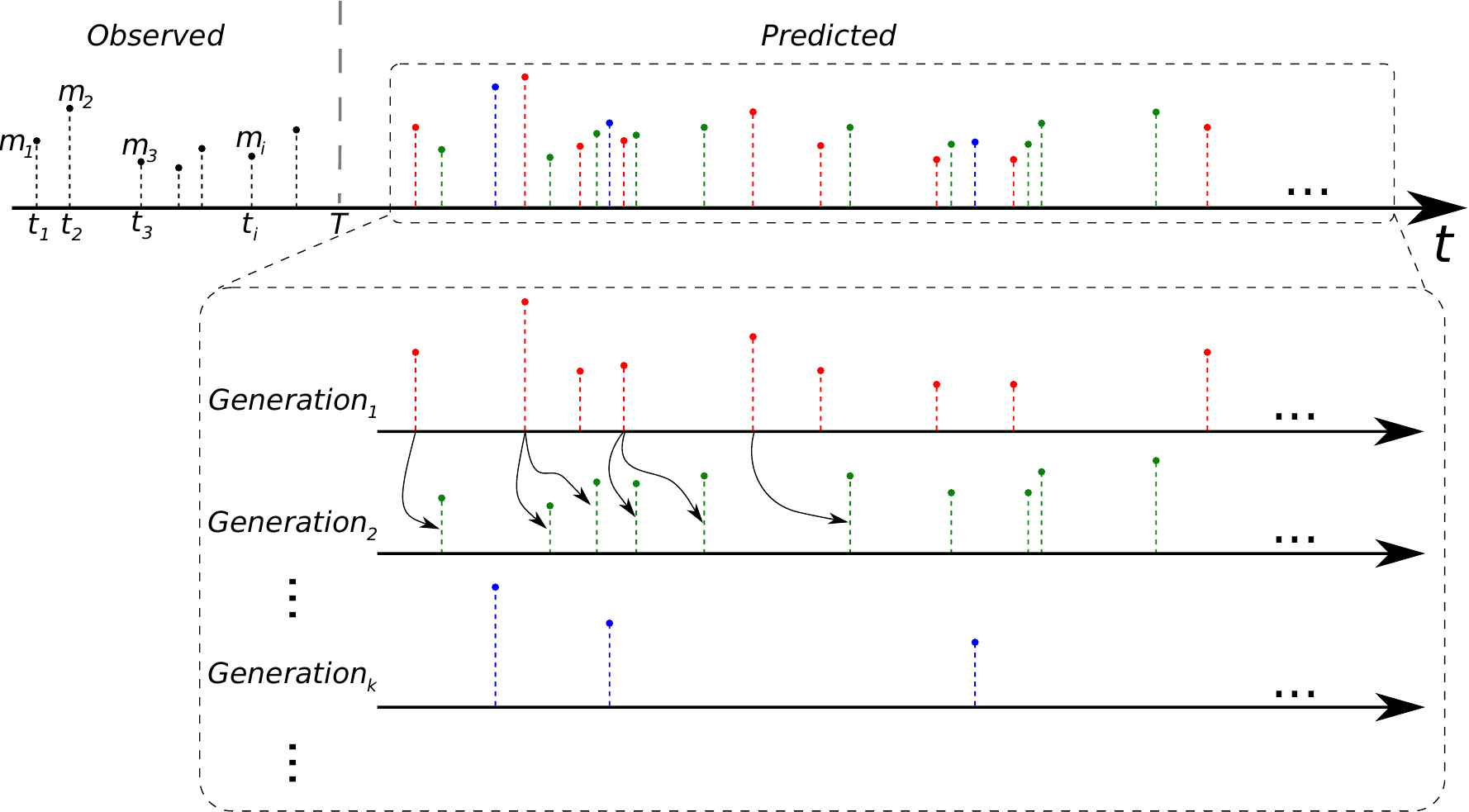}
\caption{
	Illustration of the rationale behind popularity prediction.
	The model parameters are estimated starting from a series of observed events $(m_i, t_i)$.
	Part of one possible unfolding of the diffusion cascade is simulated, using the event rate defined by Eq.~\eqref{eq:phi-separable} and 
	simulated by thinning~\cite{Ogata1999}.
	The event generations are shown: 
	events in $Generation_1$ are shown in \textcolor{red}{red} color, 
	$Generation_2$ in \textcolor{ForestGreen}{green} and
	$Generation_k$ in \textcolor{blue}{blue}.
	Note there is no theoretical limit to the number of generations or to the extent of time until the cascade dies out.
	Some of the parent-child relations between events in consecutive generations are shown.
}
\label{fig:generational-view}
\captionmoveup
\squeezeup
\end{figure}

Having observed a retweet cascade until time $T$ and fitted the parameters of the point process, one can simulate a possible continuation of the cascade, until it dies out -- assuming $n^{\ast} < 1$.
In this section, we derive the expected number of events in the cascade, over all possible continuations.
We group events based on the generation of their parents (i.e. the preceding event they are triggered by), and estimate the number of events in each generation.
%
We denote as $Generation_1$ the set of simulated events spawned by the observed events (shown in \textcolor{red}{red} in Fig.~\ref{fig:generational-view}).
Similarly, $Generation_2$ (in \textcolor{ForestGreen}{green} in Fig.~\ref{fig:generational-view}) groups events spawned by events in $Generation_1$, and so on.
Let $A_i$ be the expected number of events in $Generation_i$.
The expected number of total events in the cascade, $N_{\infty}$, is defined as:
\begin{equation}\label{eq:inf}
	\eqmoveup
	N_{\infty} = n + \sum_{i=1}^{\infty} A_i \enspace,
\end{equation}
where $n$ is the number of observed events.
For generations after the first one, the best estimate about $A_i, i \ge 2$ is constructed using the average number of children events $n^{\ast}$ and the number of events in the previous generation, i.e. $A_i = A_{i-1} n^*$.
Assuming $A_1$ known, we derive:
\begin{equation} \label{eq:gp}
	\eqmoveup
	A_i = A_{i-1} \: n^\ast = A_{i-2} \: \left( n^\ast \right)^2 = \ldots\ = A_1 \:  \left( n^\ast \right)^{i-1} , i > 1
\end{equation}
Therefore, the second term in~\eqref{eq:inf} is the sum of a converging geometric progression (assuming $n^{\ast} < 1$):
\begin{equation} \label{eq:sumgp}
	\eqmoveup
	\sum_{i=1}^{\infty} A_i = \dfrac{A_1}{1-n^\ast} \textrm{ where } n^\ast < 1
\end{equation}

$A_1$ could also be calculated in a fashion similar to Eq.~\eqref{eq:gp}.
However, a more precise estimation can be obtained, given that magnitudes of the observed events -- parents for events in $Generation_1$ -- are known:
\begin{align} \label{eq:A1}
	\eqmoveup
	A_1 &= \int_T^{\infty} \lambda(\tau)\mathrm{d}\tau = \int_T^{\infty} \sum_{t>t_i} \phi_{m_i}\left(t-t_i\right) \mathrm{d}t \nonumber \\
&= {\kappa} \sum_{i=1}^n \dfrac{{m_i}^{\beta}}{\theta \left(T+c-t_i\right)^{\theta}}
\end{align}
We obtain the estimate of the total number of events in the cascade by introducing~\eqref{eq:A1} and~\eqref{eq:sumgp} into~\eqref{eq:inf}:
\begin{equation} \label{eq:Ninf}
	\eqmoveup
N_{\infty} = n + \dfrac{\kappa} {(1-n^\ast)} \left( \sum_{i=1}^n \dfrac{{m_i}^{\beta}}{\theta \left(T+c-t_i\right)^{\theta}}\right), n^\ast < 1
\end{equation}

\subsection{Predicting total popularity}
\label{ssec:rf}

Using generative models for prediction is often sub-optimal, 
often due to simplifying assumptions of the generative process. 
Point processes in general, are optimized for explaining observed event history and not optimized for prediction. 
Zhao \etal~\cite{Zhao2015}, for example, recently deployed a number of heuristic correction factors to discount the initial burst and account for a long-term decay.
We consider that the Hawkes point process can benefit from systematically 
learning a predictor, to better account for data noise and limiting model assumptions.
First, it is well know that using the number of followers as user influences $m_i$ is 
at best an approximation~\cite{cha2010million} and often does not scale with retweeting propensity. 
Second, the network in which cascades happen can be inhomogenous over the lifetime 
of the diffusion, e.g. users who respond early may have a inherently shorter memory than those responding late. 
Last, point processes cannot generate a meaningful prediction for supercritical cascades, 
and the model estimation can be prone to local minima. 
In this section, we leverage a collection of previously observed cascade histories to fine tune popularity prediction. 
This predictive layer accounts for the limitations above, and places point process estimates in a comparable 
framework as the feature-driven approaches. 


\textbf{Prediction setup.}
Each retweet cascade is described using four features $\{ c, \theta, A_1, n^\ast\}$.
These four features were found to be the most correlated with $\epsilon = \left( N_{real} - N_\infty \right)^2$, the error made when predicting popularity using $N_\infty$.
Furthermore, $n^\ast$ and $\theta$ are expressed as percentiles over the range observed in the training set, to account for their non-linear relation with $\epsilon$.
Note that parameters $\kappa$ and $\beta$ are not used: the information in $\kappa$ is already included in $A_1$, whereas $\beta$ was found non-correlated with $\epsilon$.

In the classification task detailed in Sec.~\ref{subsec:classification-task}, the Random Forest is trained to output the selected class.
In the popularity prediction task -- Sec.~\ref{subsec:compare-generative} and~\ref{subsec:compare-data-driven} -- we train the Random Forest to output a scaling factor $\omega$ for the expected number of future events in a given retweet cascade.
The final corrected popularity prediction is computed as:
\begin{equation} \label{eq:Nhatinf}
	\eqmoveup
	\hat{N}_{\infty} = n + \omega \left(\dfrac{A_1}{1-n^*}\right) \enspace.
\end{equation}

\begin{figure}[tbp]
    \centering
    \subfloat[] {
		\includegraphics[height=0.123\textheight]{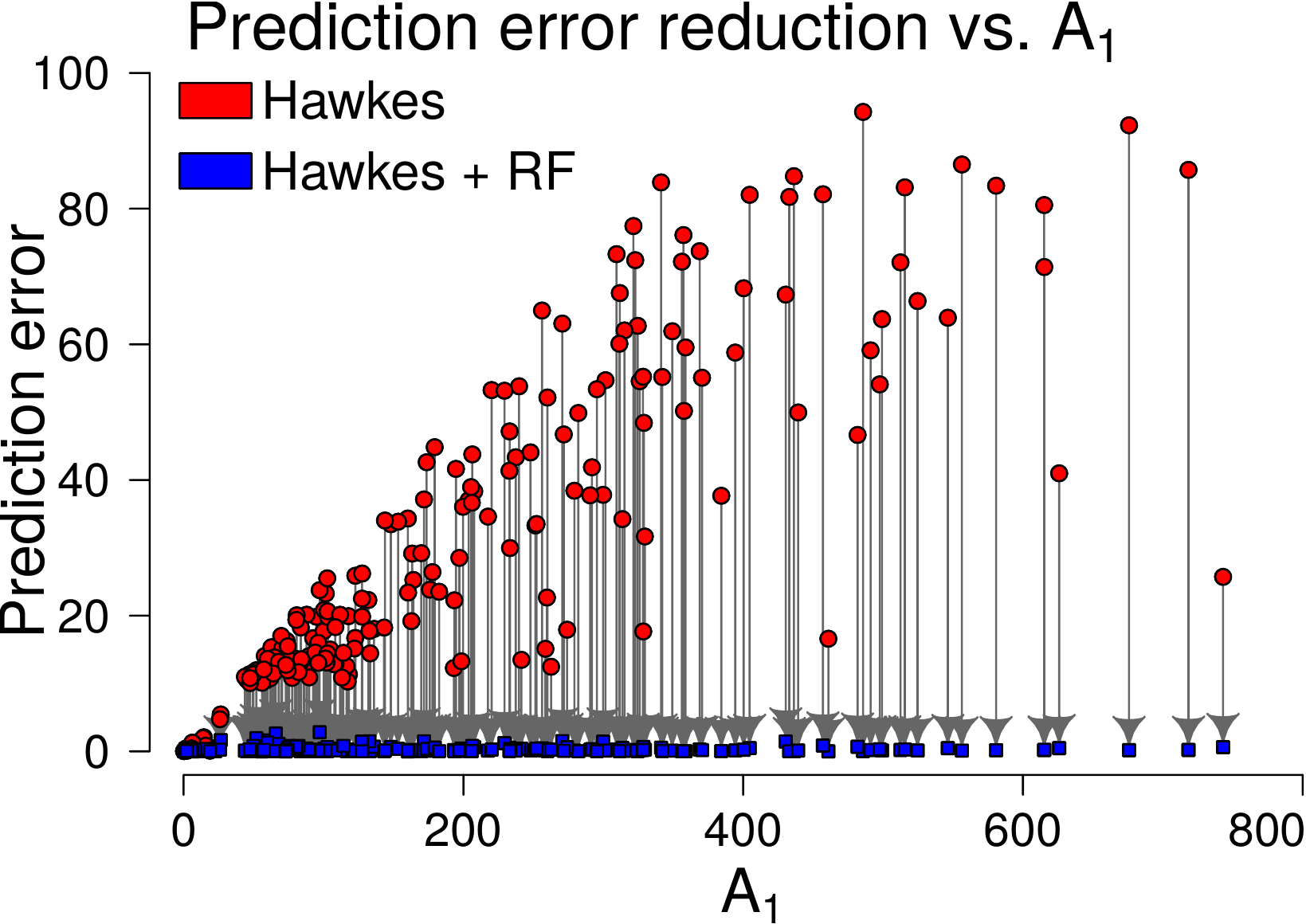}
	} 
	\subfloat[] {
		\includegraphics[height=0.123\textheight]{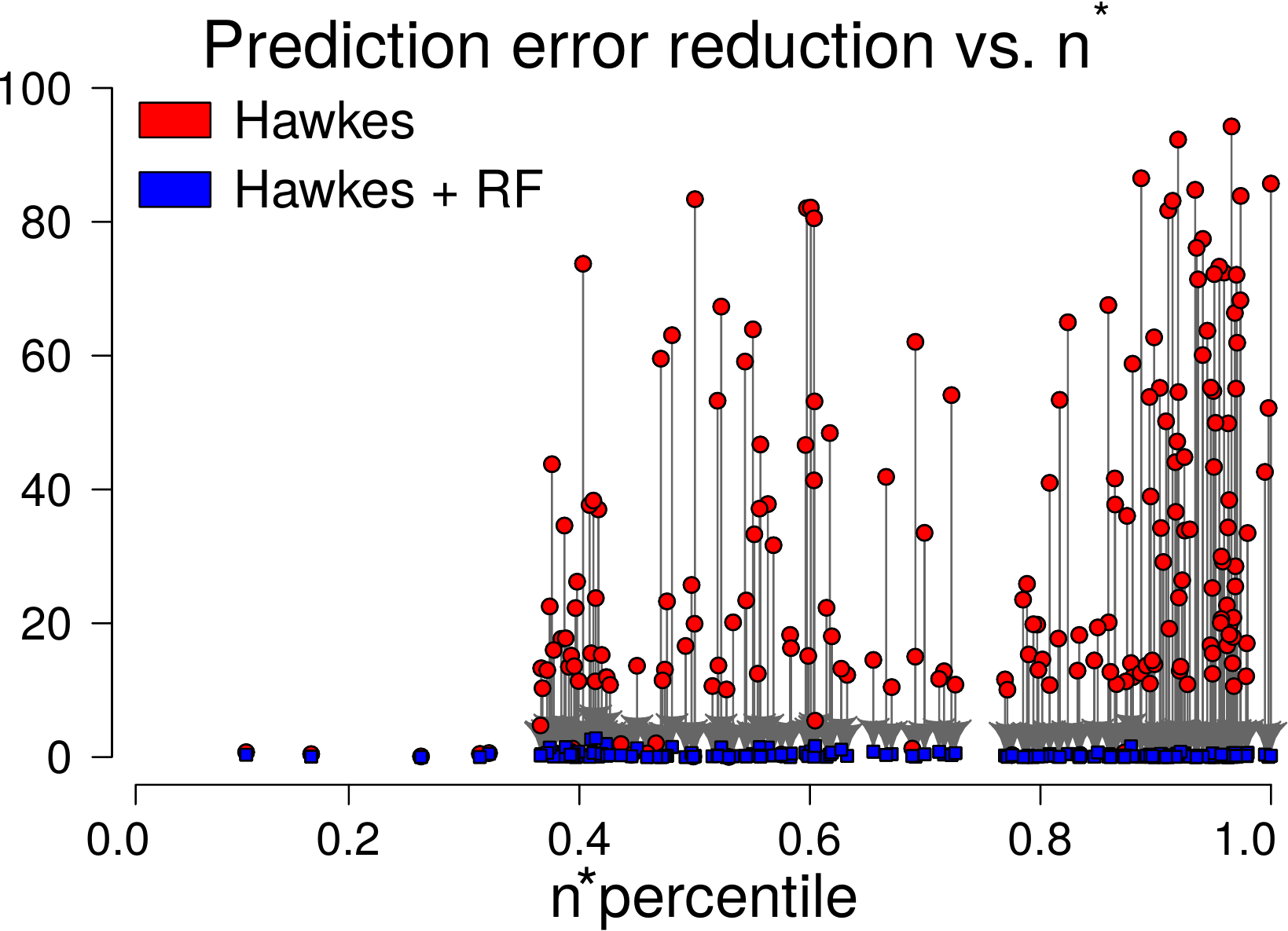}
	}
    \caption{
	Reduction of prediction error for a subset of cascades.
	For each cascade, the error made when predicting final popularity using the theoretical $N_\infty$ is shown using \textcolor{red}{red} circles, and with  \textcolor{blue}{blue} squares the error made when using the predictive layer $\hat N_\infty$.
	Each gray arrow shows the reduction of error for a single cascade, and pairs together a red circle with a blue square.
	The error reduction is show in relation to $A_1$ (a) and $n^\ast$ (b).
    }
    \label{fig:effect-predictive-layer}
    \captionmoveup
\squeezeup
\end{figure}

\textbf{Effects of the predictive layer.}
The immediate effect of introducing the predictive layer is the reduction of the ARE prediction error, measured as detailed in Sec.~\ref{subsec:compare-generative}.
Fig~\ref{fig:effect-predictive-layer} illustrates this reduction on a subset of the cascades in the \news dataset -- described in Sec.~\ref{subsec:dataset} -- and its relation with $A_1$ and $n^\ast$.
The length of the arrows is proportional with the error reduction.
The range the errors made without the predictive layer increases linearly with $A_1$.
Consequently the error reduction when using $\hat N_\infty$ is generally higher when $A_1$ is higher (Fig.~\ref{fig:effect-predictive-layer}a).
The relation between $n^\ast$ and the prediction error is non-linear, therefore in  Fig.~\ref{fig:effect-predictive-layer}b the horizontal axis shows the percentile values of $n^\ast$.
The variation of the error reduction has two maxima, one around 40\% and another one around 95\%.
Fig.~\ref{fig:effect-predictive-layer} can be summarized as follows: 
the error reduction increases linearly with the expected number of events in $Generation_1$, but exhibits two maxima in relation with the expected number of events in $Generation_i, i \geq 2$.
Another interesting observation is that $N_\infty$ tends to over-estimate cascade sizes, in other words cascades tend to achieve less popularity than expected when observed in isolation.
This aligns with the hypothesis that content items compete with each other for a finite amount of human attention~\cite{MIR13}.


\section{Data and features for prediction}
\label{sec:data}

In this section we describe the two datasets, \seismicdata and \news, used in our experiments. 
Next, we describe the set of features constructed for our feature-driven approach.

\subsection{Datasets}
\label{subsec:dataset}

\seismicdata: 
Zhao \etal~\cite{Zhao2015} made publicly available the dataset they used to predict retweet cascade sizes.
It contains a sample of all tweets during a month (i.e. using the \emph{firehose} Twitter API restricted access), further filtered so that the length of each cascade is greater than 50. 
For each cascade, the dataset contains information about the original tweet id and start time and, for each of the retweets in the cascade, the time offset of the retweet and the number of followers of the user posting the retweet. 
We randomly sample the cascades in the original dataset and we construct \seismicdata to contain 30,463 cascades of length between 50 and 5000. 
The resulting dataset has a mean cascade length of 160 and a median of 95. 

\textbf{\news:} 
We construct a domain specific dataset of tweets dealing with news articles, by collecting data using the \emph{free} Twitter Streaming API\footnote{\url{https://dev.twitter.com/streaming/overview}}, over a period of four moths from April 2015 to July 2015.
We track the official twitter handles of ten popular news outlets, namely NewYork Times, Associated Press, CNN, Washington Post, Reuters, Yahoo News, The Guardian, BBC, Huffington Post and Google News. 
We also capture all tweets containing either i) user mentions of the above users or ii) a commonly used shortened url towards any of the ten news websites. 
This leads to a collection of 49,735,271 tweets.
Most of the cascades in the \news dataset are of length one or two and, for compatibility with the \seismicdata, we present the statistics for cascades of length over 50:
110,954 cascades, mean cascade length is 158 and median length is 90.

\textbf{Difference between datasets.}
The \news dataset contains several additional information, when compared to \seismicdata:
for each user in a retweet cascade, we record her number of friends, number of posted statuses and the account creation time (these information are recorded in every tweet we capture through the API).
This additional information is required by the feature-driven method described in Section~\ref{subsec:data-driven}, therefore in the experiments in Sec.~\ref{sec:xp}, feature-based methods cannot run on \seismicdata.

\textbf{An open-access, domain-specific dataset.}
Many of the datasets used by prior work for popularity prediction were not made public because of user agreement terms and they are difficult to replicate.
The barriers for having an open benchmark include
proprietary insider information (e.g. complete Facebook network measurements~\cite{CHE14}), privileged API access (e.g., Twitter firehose~\cite{Zhao2015}) or contain restricted data from non-English sources~\cite{Ding2015,yu2015micro}.
Furthermore, we argue that predicting popularity on a topic- or domain- specific dataset 
is as relevant as that on the firehose feed -- content producers and consumers
are often interested in content within a source (e.g. NYTimes) or a topic (e.g. technology).
Therefore, we chose to construct \news, our domain-specific dataset, using only free access APIs. 
We present a strong basis for future benchmarking 
by combining the descriptive power of \featuredriven and modeling capabilities of \hawkes in Sec.~\ref{sec:xp}.

\subsection{Features for popularity prediction}
\label{subsec:data-driven}

Our feature-driven predictor describes each cascade using four types of features:
a) basic user features, b) temporal features, c) volume features and d) past user success.
They are chosen as being the
most informative in recent literature~\cite{BAK11,CHE14,Martin2016,Pinto2013,szabo2010} 
and require only free access to the Twitter data source.
In particular, Martin \etal~\cite{Martin2016} compare a wide range of features, including user, content and user-content interplay features, and they show that basic user features together with past user success features account for $87\%$ of the prediction performance.
Cheng \etal~\cite{CHE14} showed that temporal features relating to the unfolding of the observed part of the diffusion account for the bulk of the total classification performances.
They report only 0.025 additional accuracy when using all the other features combined in addition to temporal features.
Lastly, Szabo \etal~\cite{szabo2010} and later Pinto \etal~\cite{Pinto2013} show that early popularity is predictive for total popularity.

\textbf{Basic User Features}~\cite{BAK11,CHE14,Martin2016}.
Basic user statistics capture the social influence of a user.
When observing the prefix of a cascade, we use the five point summary (min, median, max, 25-th and 75-th percentile) to represent the distribution for a feature:
\begin{itemize}
\setlength{\parskip}{-2pt}
\item Number of Followers: count of user followers;
\item Number of Friends: count of friends for a user;
\item Number of Status: count of statuses posted by user;
\item User Time: Time when user account was created.
\end{itemize}

\textbf{Temporal Features}~\cite{CHE14}. 
We capture the temporal dynamics of a cascade using the following features:
\begin{itemize}
\setlength{\parskip}{-2pt}
\item First Half Rate: mean waiting time between retweets, during the first half of the observed cascade;
\item Second Half Rate: mean waiting time between retweets, during the second half of the observed cascade;
\item Waiting Time Distribution: five point summary of waiting times between retweets, in the observed cascade;
\item Exposure Distribution: five point summary of the distribution of number of followers for all users before the last retweet, in the observed cascade. Used only in the classification task in Sec.~\ref{subsec:classification-task}.
\end{itemize}

\textbf{Volume}~\cite{Pinto2013,szabo2010}.
Number of retweets in the observed part of the cascade, which captures early popularity. 

\textbf{Past User Success}~\cite{BAK11,Martin2016}. 
Average size of the cascades started by a given user in past. 
For an observed cascade, we use the five point summary of the distribution of past user success for all previously known users who participate in the cascade.
This feature requires a large volume of historical data and can be computed only for users that have started cascades in the past.
We detail its construction in Sec.~\ref{subsec:compare-generative}.


\section{Experiments}
\label{sec:xp}

We first present in Sec.~\ref{subsec:compare-generative} a thorough comparison between the Hawkes model and \seismic, the state-of-the-art generative model on popularity prediction, using two datasets.
In Sec.~\ref{subsec:compare-data-driven} and Sec.~\ref{subsec:classification-task} we compare the feature-driven classifier to generative approaches. 
Sec.~\ref{subsec:compare-generative} and Sec.~\ref{subsec:compare-data-driven} tackle regression tasks, i.e., predicting total cascade size after observing it for a certain time, while Sec.~\ref{subsec:classification-task} tackles classification tasks, i.e., whether or not the cascade size will double from what has been observed. 

\subsection{Cascade size: two generative methods}
\label{subsec:compare-generative}

\begin{table*}[tbp]
\caption{
	(left hand side) Mean Average Relative Error (ARE) $\pm$ standard deviation obtained using \hawkes and \seismic on the \seismicdata and \news datasets. 
	(right hand side) Number of cascades for which prediction failed.}
\centering
\begin{tabular}{ccccc|ccc}
\toprule
& & \multicolumn{3}{ c| }{Prediction error: different time lengths} & \multicolumn{3}{ c }{Number of failed cascades}\\ 
Dataset & Approach & 5 minutes & 10 minutes & 1 hour & 5 minutes & 10 minutes & 1 hour \\ \midrule
\multirow{2}{*}{\seismicdata}  & \seismic & 2.61 $\pm 55.80$ & 0.70 $\pm 15.58$ & 0.51 $\pm 10.81$ & 507 & 164 & 71 \\
& \hawkes & 0.36 $\pm 0.52$ & 0.33 $\pm0.41$ & 0.30 $\pm 0.38$ & 302 & 105 & 58 \\ \midrule
\multirow{2}{*}{\news} & \seismic & 11.13 $\pm 282.96$ & 0.84 $\pm 11.77$ & 0.33 $\pm 0.92$ & 1022 & 155 & 123 \\ 
& \hawkes & 0.42 $\pm 6.83$ & 0.25 $\pm 0.60$ & 0.22 $\pm 1.16$ & 149 & 45 & 37 \\
 \bottomrule
\end{tabular}
\label{tab:generative}
\end{table*}

\begin{figure}[tb]
    \centering
	\includegraphics[width=0.147\textwidth]{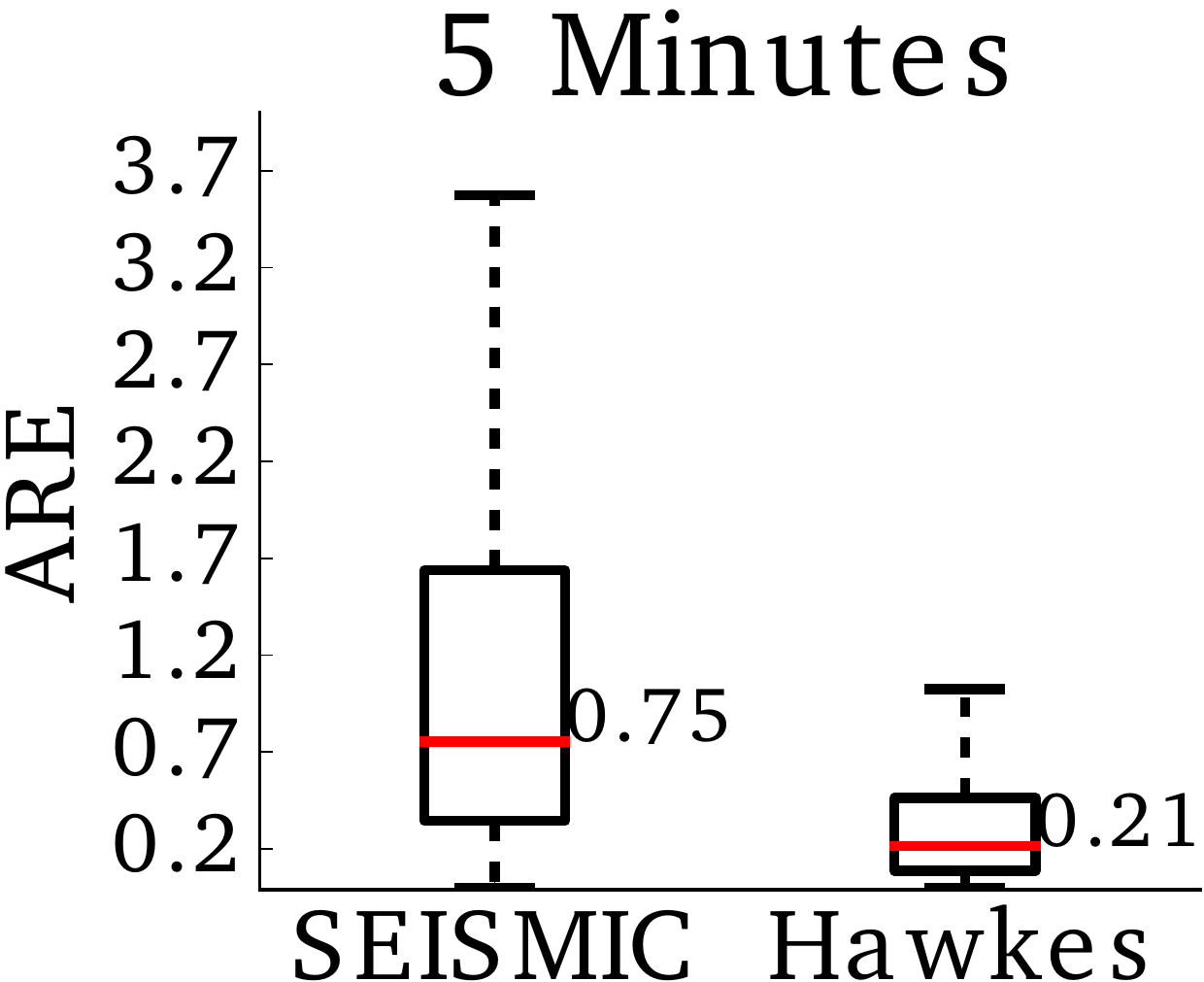}
	\includegraphics[width=0.147\textwidth]{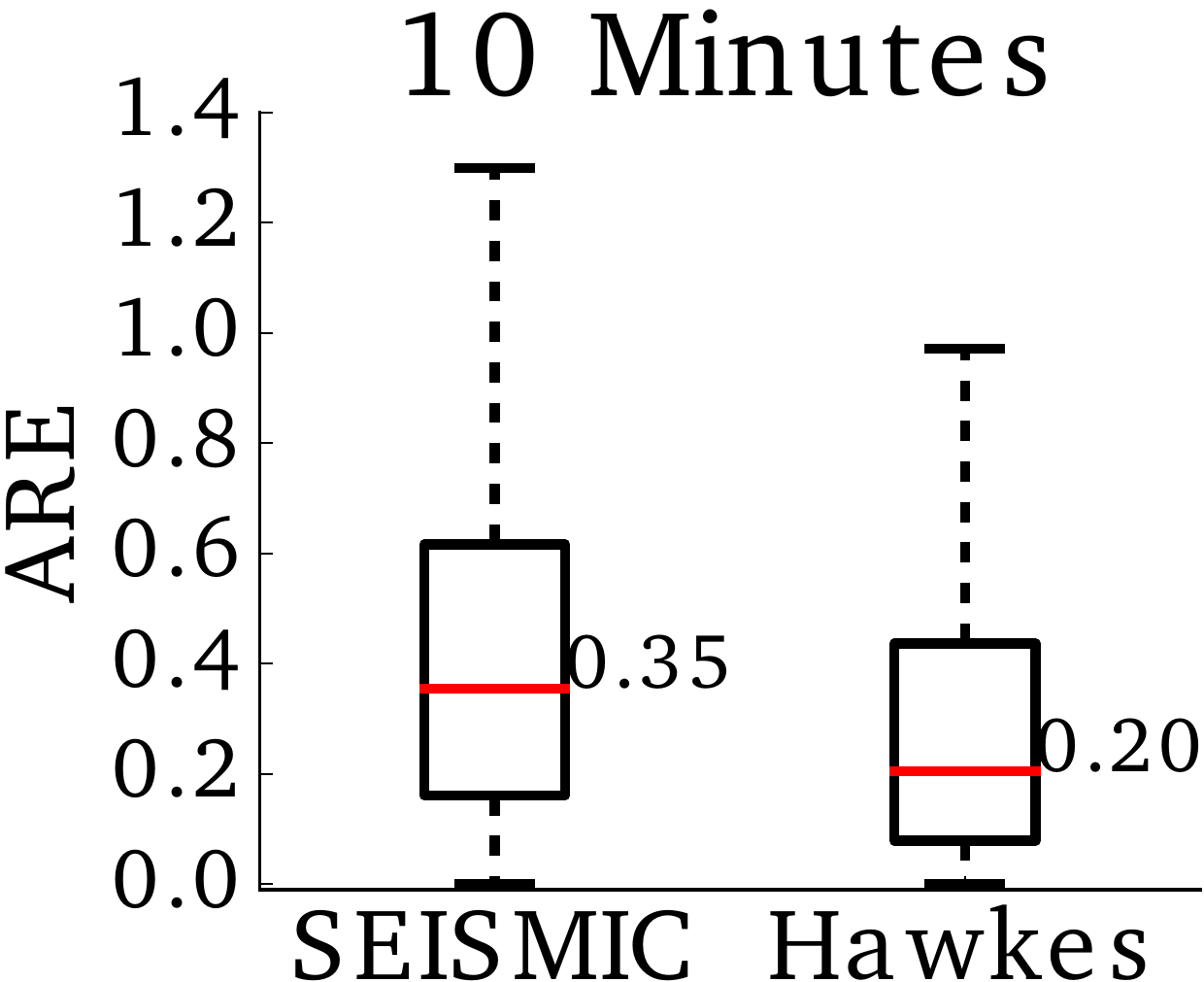}
	\includegraphics[width=0.147\textwidth]{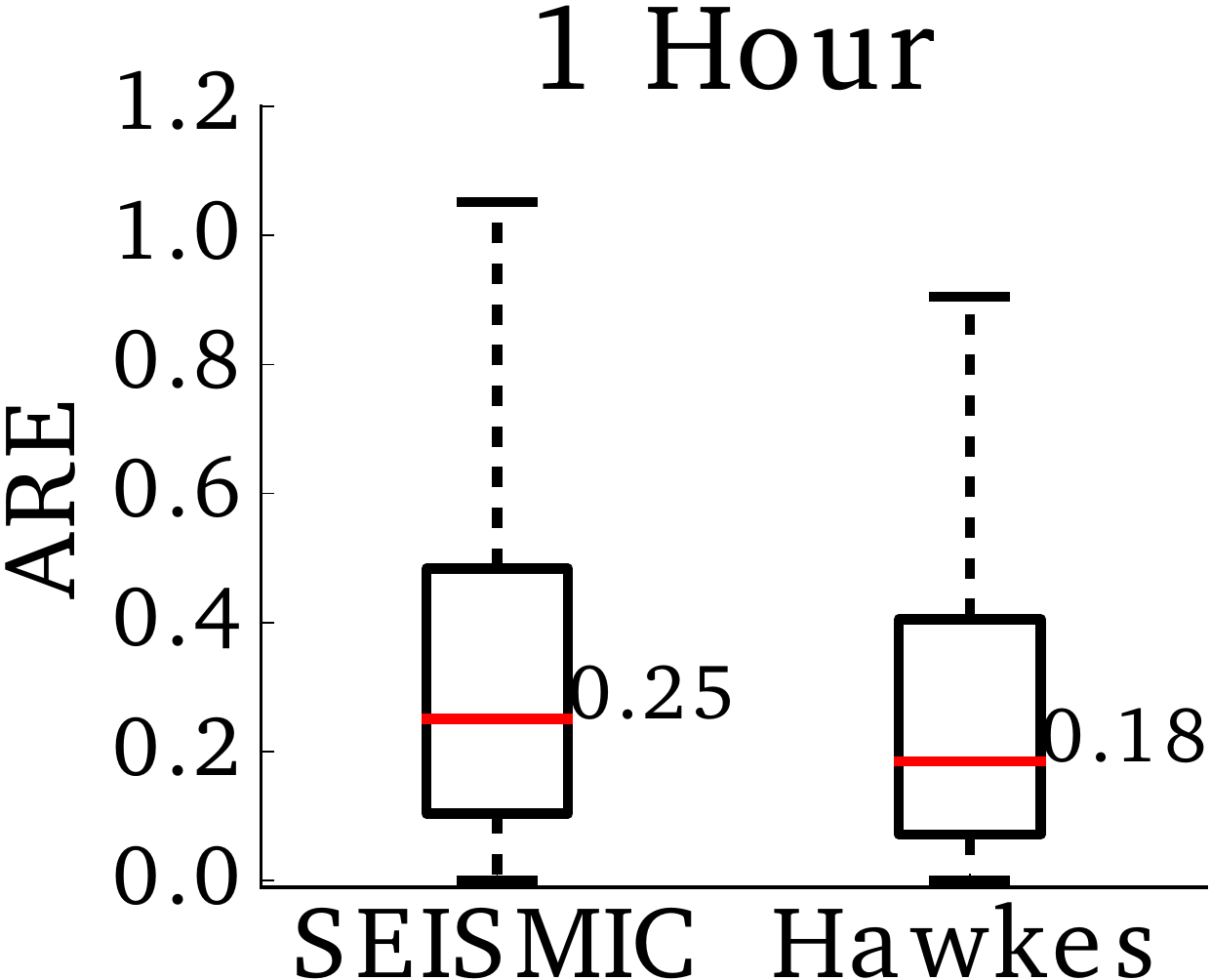}
	\vspace{0.1cm} \\
	\includegraphics[width=0.147\textwidth]{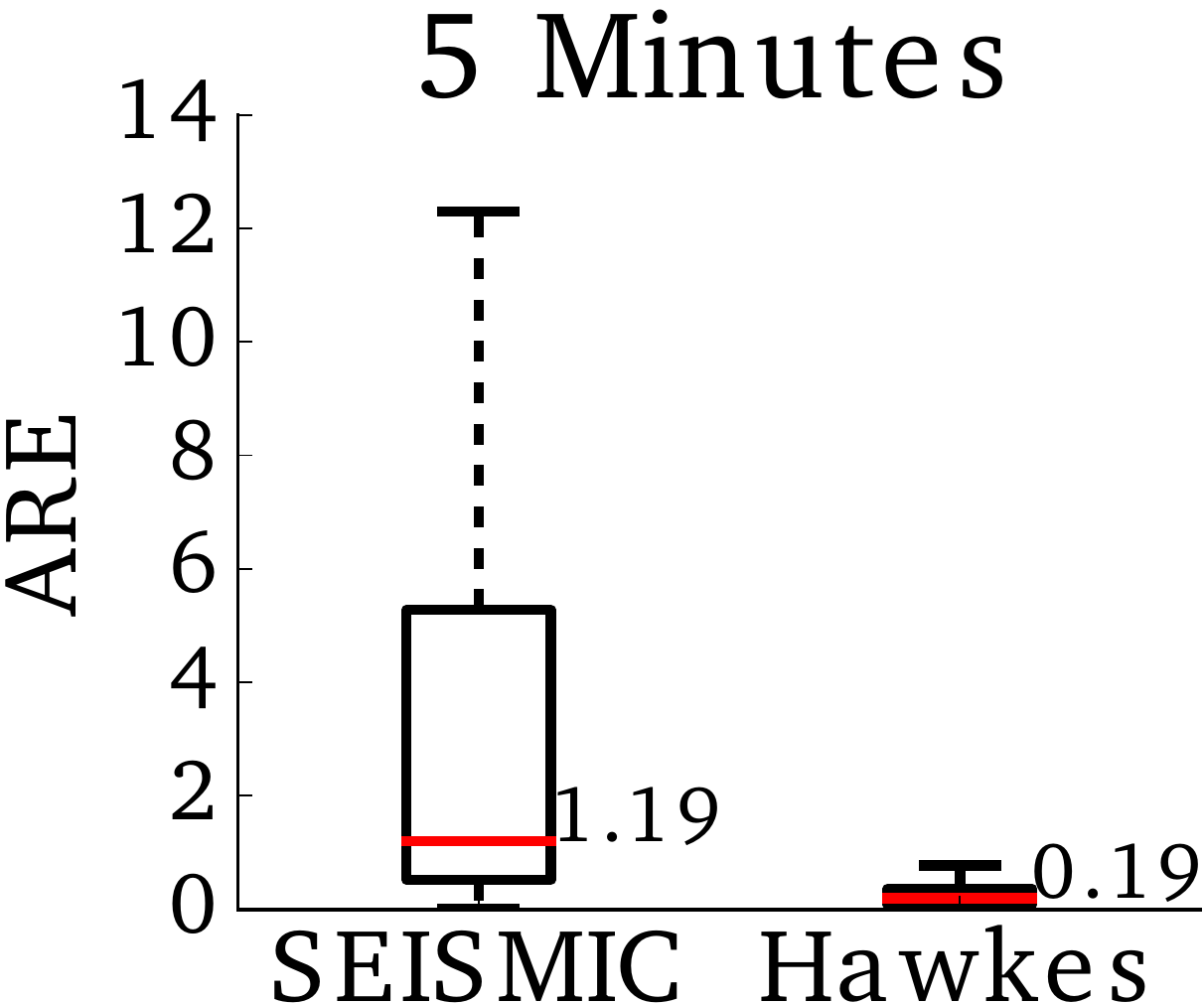}
	\includegraphics[width=0.147\textwidth]{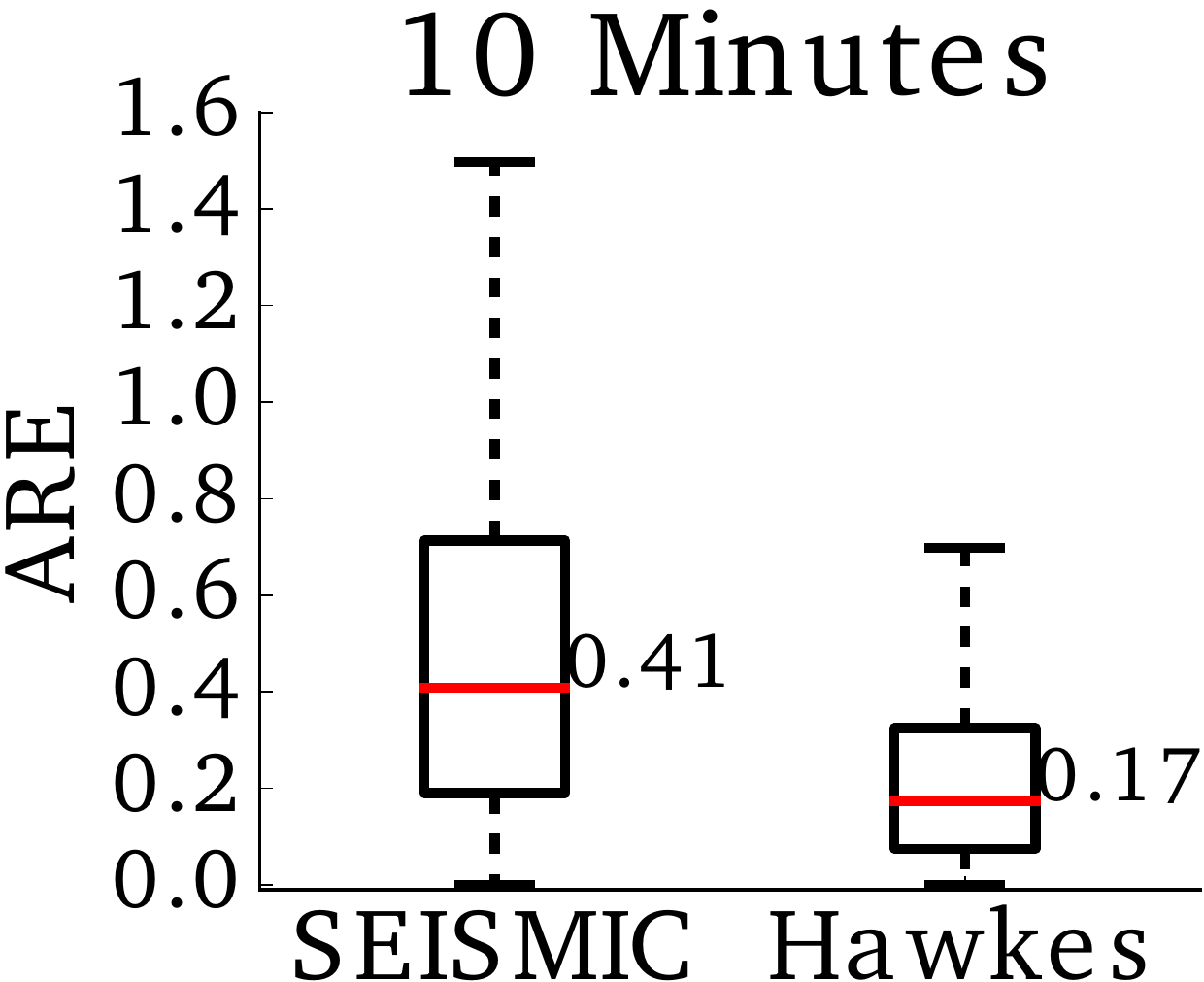}
	\includegraphics[width=0.147\textwidth]{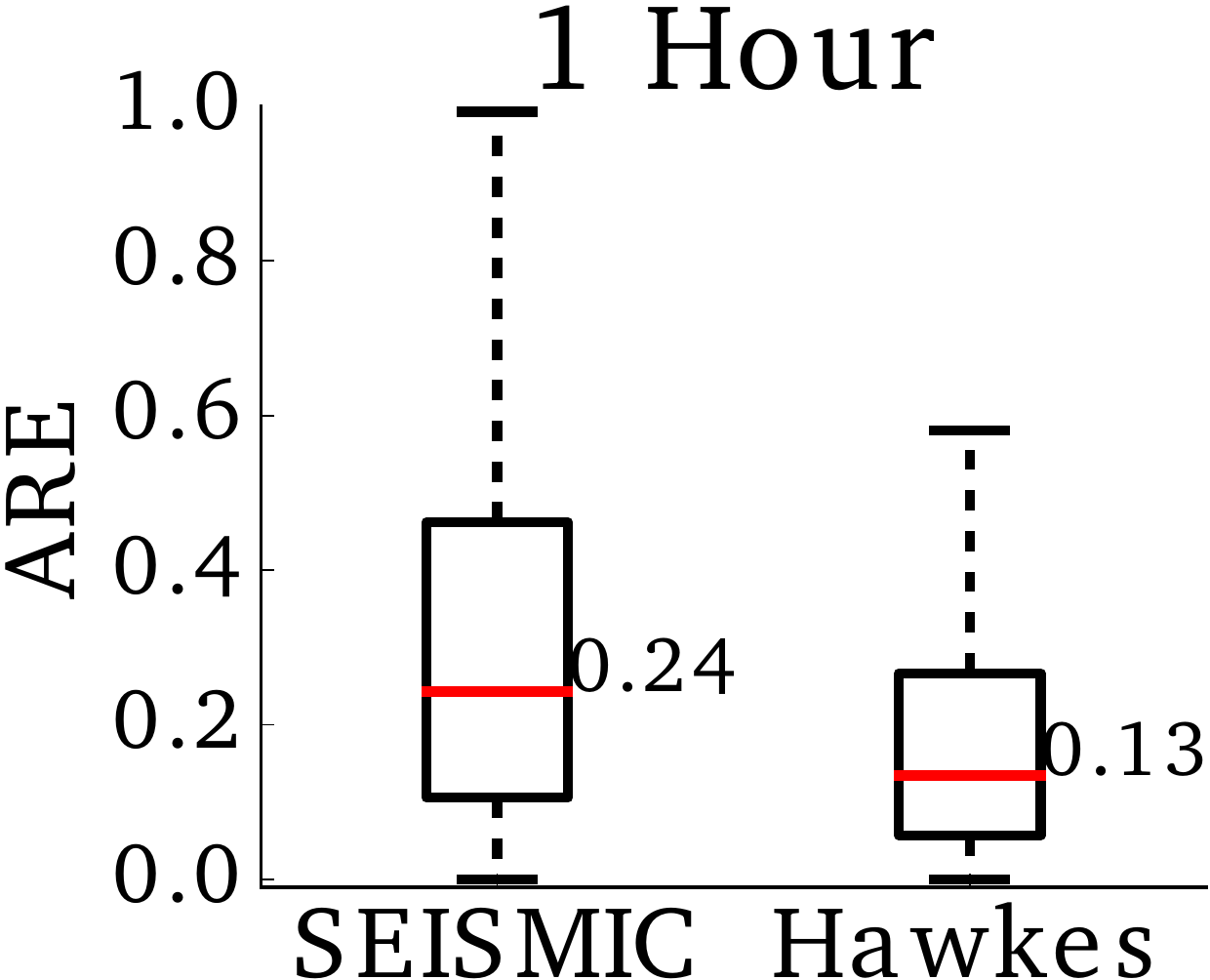}
    \caption{
    Distribution of Absolute Relative Error (ARE) on the \seismicdata dataset (top row) and on the \news (bottom row), for \seismic and \hawkes.
    A part of the diffusion cascade is observed before making the predictions: 5 min (left column), 10 min (middle column) and 1 hour (right column).
    The \textcolor{red}{red} line and the numeric annotations denotes the median values of the distributions.
    }
    \label{fig:gen-news-comparision}
    \captionmoveup
\end{figure}
We compare our point processes model, \hawkes, with \seismic on two datasets, \seismicdata and \news.
Note that \seismicdata is the dataset on which Zhao \etal~\cite{Zhao2015} reported results, while \news dataset supports extracting a richer set of features that is not available in \seismicdata.
In addition, Zhao \etal~\cite{Zhao2015} showed that \seismic compared favorably against a number of baselines: from auto-regressive and  generative models such as dynamic Poisson model~\cite{Agarwal2009,Crane2008} and reinforced Poisson model~\cite{Shen2014}, to linear regression and it's variants~\cite{szabo2010}. 

\textbf{Experimental setup.}
We run our experiments on the \seismicdata dataset and a subset of \news: 20,093 cascades from the month of July, which have the total length of at least 50.
Both algorithms observe the initial part of each cascade for a limited extent of time and predict its final popularity. We estimate separate sets of parameters for \hawkes and \seismic 
after observing the cascades for 5 minutes, 10 minutes and 1 hour, respectively.
After the fitting is finished, we train the Random Forest regressor in the predictive layer of the \hawkes model.  Performance of this regression layer is reported using ten fold cross-validation. 40\% of cascades are used for training, 60\% for testing. 
All experimental protocol choices -- minimum cascade length threshold, random training/testing split and length of time prefixes -- are made to mimic closely the original experimental setup~\cite{Zhao2015}. 

%
We use the Absolute Relative Error (ARE) to measure prediction performance on a given cascade $w$:
\begin{equation*} \label{eq:ARE}
	ARE^w = \dfrac{\left|\hat{N}_{\infty}^w - N_{real}^w\right|}{N_{real}^w} \enspace,
\end{equation*}
where $ \hat{N}_{\infty}$  and $N_{real}$ are the predicted and observed popularities of cascade $w$.

\begin{table}[tbp]
\caption{
	Mean ARE $\pm$ standard deviation for \seismic, \hawkes, \featuredriven and \hybrid models for the \news July'15 dataset.
}
\begin{tabular}{cccc} 
\toprule
Approach & 5 minutes & 10 minutes & 1 hour \\ 
\midrule
\seismic & 15.16 $\pm 375.08$ & 0.71 $\pm 4.89$ & 0.32 $\pm 0.40$ \\
{\sc Feat.-Driven} & 0.25 $\pm 0.18$ & 0.22 $\pm 0.17$ & 0.17 $\pm 0.14$     \\ 
\hawkes & 0.27 $\pm 1.83$ & 0.22 $\pm 0.80$ & 0.17 $\pm 0.36$    \\
\hybrid & 0.17 $\pm 0.16$ & 0.15 $\pm 0.14$ & 0.11 $\pm 0.12$      \\ 
\bottomrule
\end{tabular}
\label{tab:feature-driven-comparison}
\end{table}

\textbf{Results.}
The prediction results are summarized in Fig.~\ref{fig:gen-news-comparision} as boxplots of ARE for each combination (dataset, approach, observed time), while Table~\ref{tab:generative} shows the mean ARE $\pm$ standard deviation. Note that generative models do not output a prediction for all cascades. 
For example, when branching factor $n^\ast \geq 1$ for \hawkes the estimate for cascade size is infinite, similar failures also occur with \seismic when the infectiousness parameter $p\geq\frac{1}{n^*}$ (\cite{Zhao2015} Eq(7)).
Here we report the number of ``failed'' cascades in Table~\ref{tab:generative}, 
and only report the average ARE on cascades that both approach produce a prediction. 
As shown in right hand side of the Table~\ref{tab:generative}, 
we observe that \hawkes is able to produce meaningful predictions for more cascades than \seismic.
For example, on \news data with 5 minutes, we output prediction for $873$ (or 0.7\%) more cascades than \seismic. 


%
We can see that \hawkes consistently outperforms \seismic, as it achieves lower average ARE on both datasets. 
Mean ARE for \hawkes is at least $40\%$ better than \seismic on \seismicdata and $33\%$ on \news.
We observe that \news dataset after 5 minutes of observation seems the most difficult to predict: \hawkes produces a mean ARE of $0.42$, while the mean ARE for \seismic is $11.13$. 
Fig.~\ref{fig:gen-news-comparision} (top row) shows that the median of \hawkes is at least $25\%$ better than \seismic for any time prefix, suggesting that our approach predicts better for most of the cascades in \seismicdata.
Similar conclusions can be drawn from the bottom row of Fig.~\ref{fig:gen-news-comparision} (\news data): the median ARE of \hawkes is at most $0.19$ at 5 minutes, while the best of \seismic models is a median ARE of $.24$ at 1 hour. 

We are glad to see that the \hawkes model achieves better predictive performance than \seismic, by exploiting full flexibility in its non-linear parameters, 
and a predictive layer that systematically optimizes for future prediction using information from other cascades. 

\subsection{Cascade size: generative vs. feature-driven}
\label{subsec:compare-data-driven}

\begin{figure}[tb]
    \centering
 	\subfloat[] {
        \includegraphics[width=0.23\textwidth]{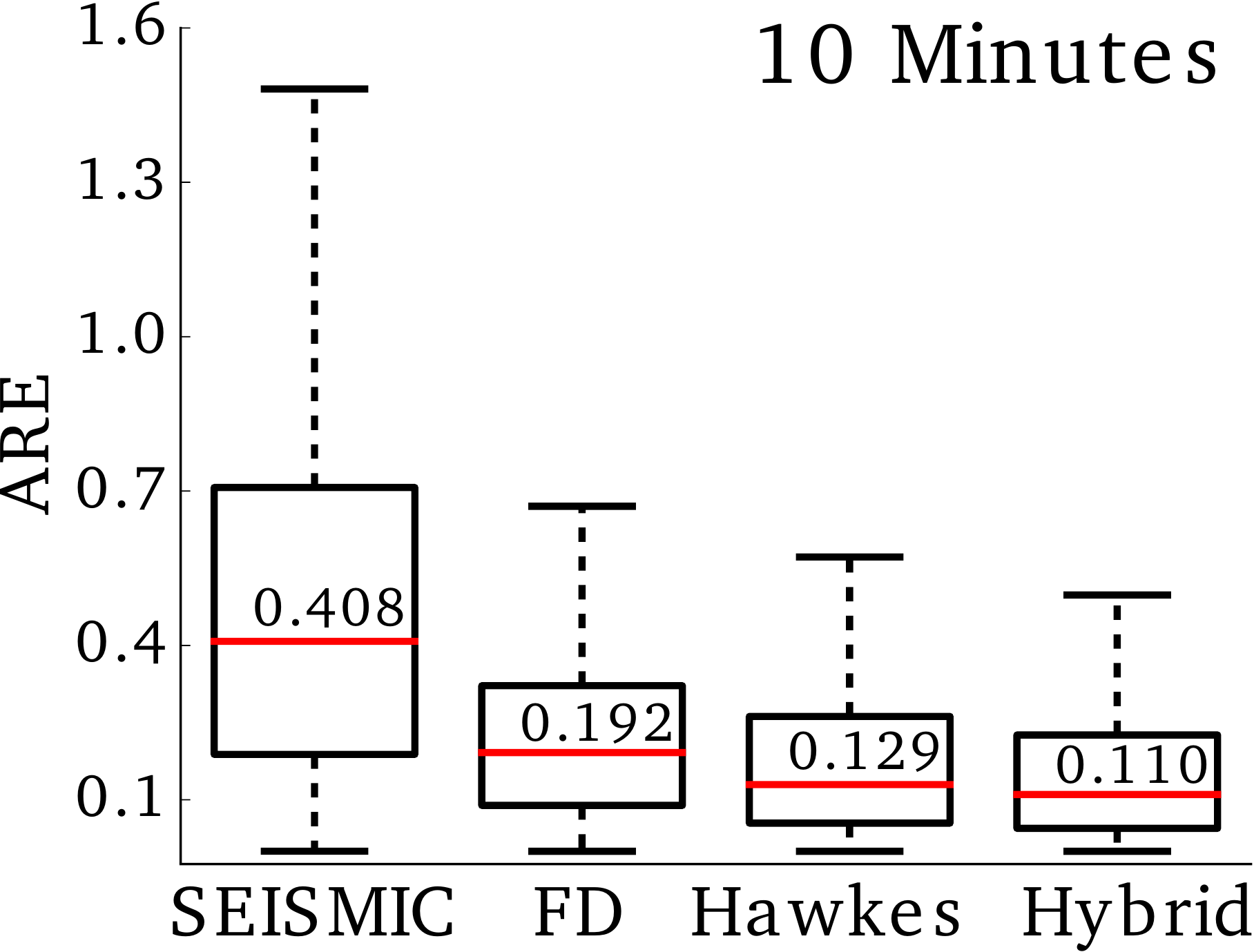}
	}
 	\subfloat[] {
        \includegraphics[width=0.23\textwidth]{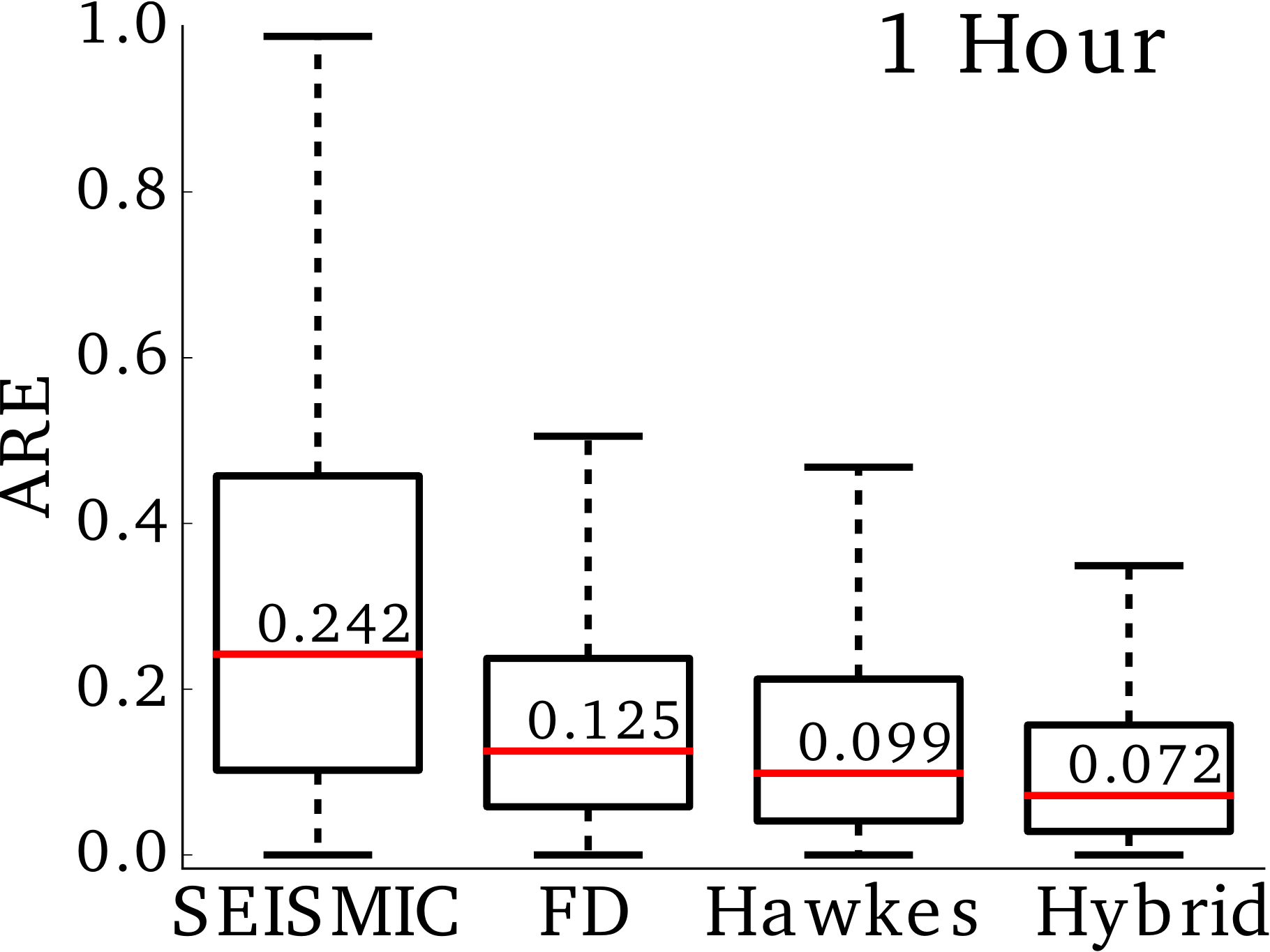}
	} 
    \caption{Distribution of ARE on the \news dataset, split in time for July, for \seismic, \featuredriven, \hawkes and \hybrid, after observing 10 minutes (a) and 1 hour (b).
    The \textcolor{red}{red} line and the numeric annotations denote median value.}
    \label{fig:hybrid-comparison}
    \captionmoveup
\end{figure}

In this section we compare the performances of feature-driven and generative approaches on the \news dataset only, as the \seismicdata does not contain the necessary data to construct the features detailed in Sec.~\ref{subsec:data-driven}.
 
\begin{figure*}[tbp]
    \centering
		\includegraphics[height=0.123\textheight]{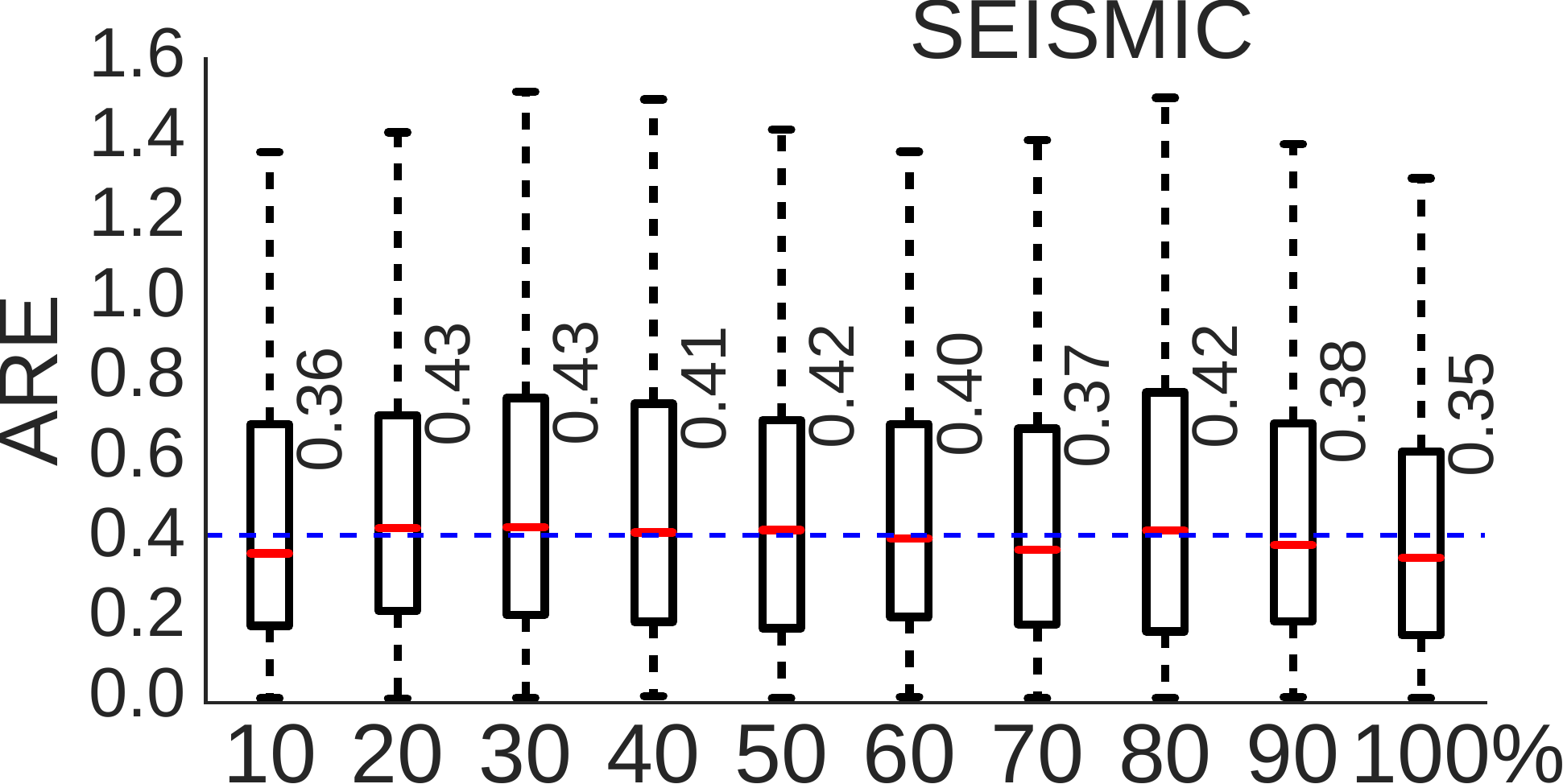}
		\includegraphics[height=0.123\textheight]{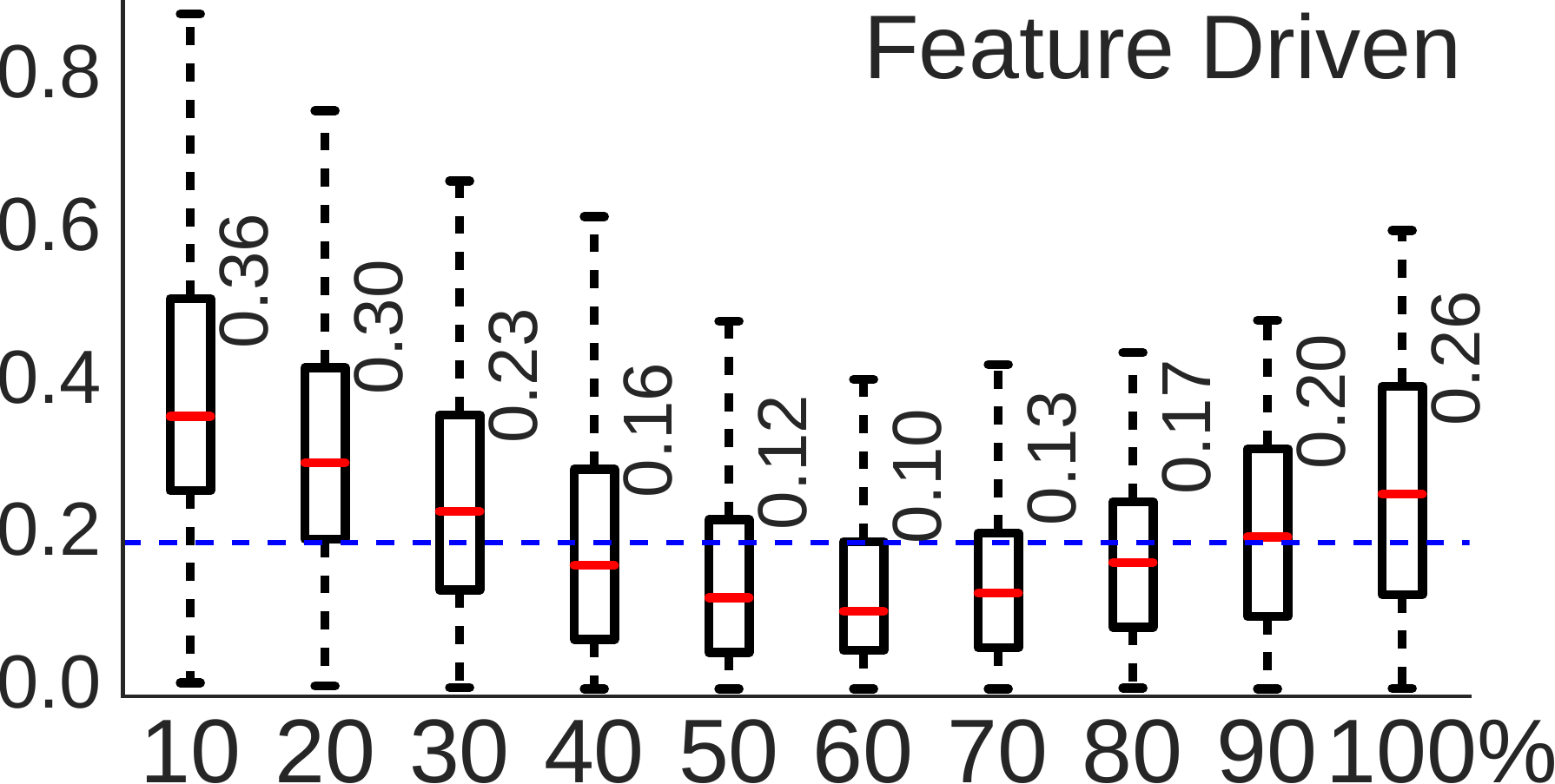}
		\includegraphics[height=0.123\textheight]{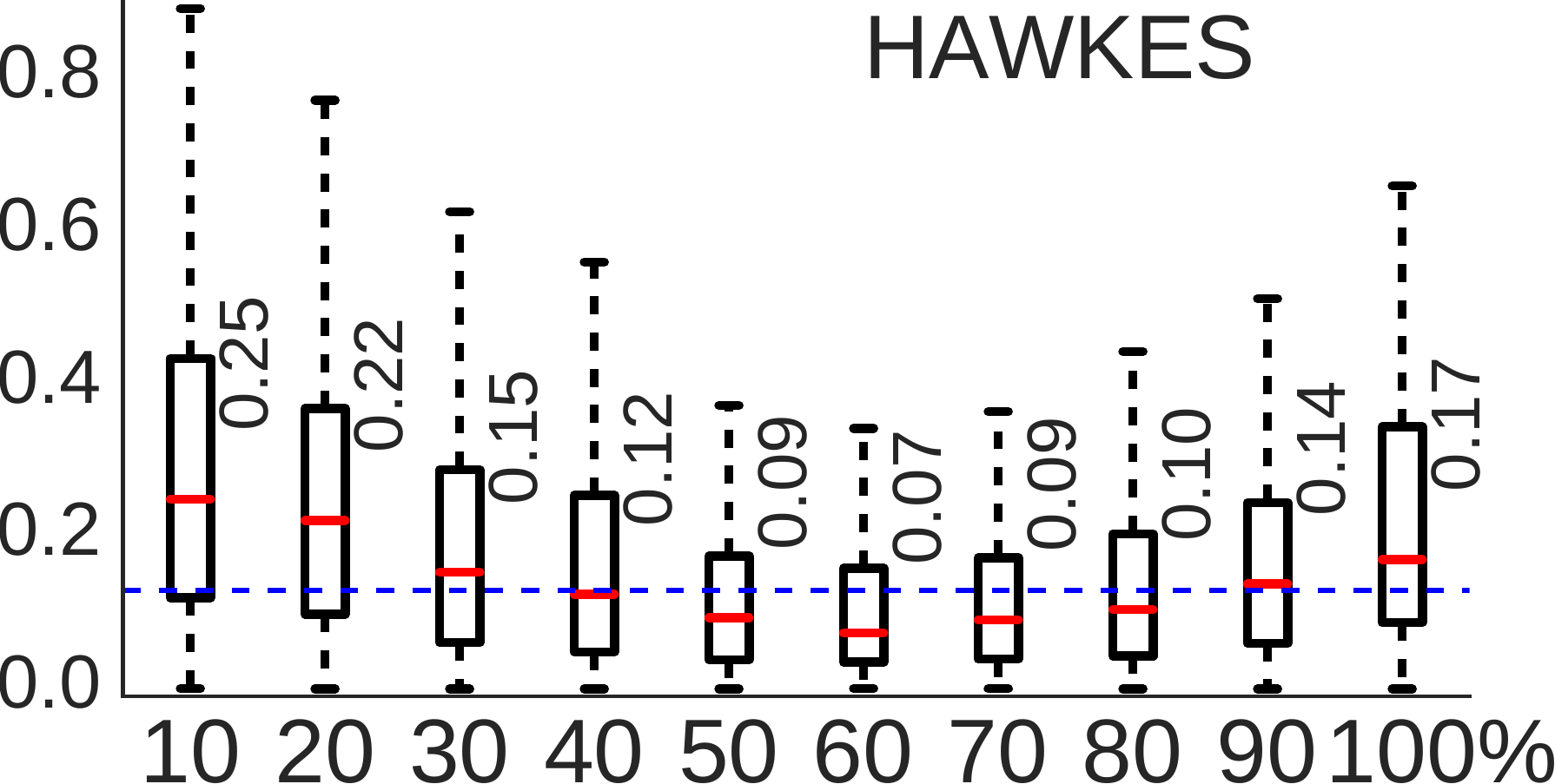}
    \caption{
	ARE distribution over popularity percentiles, on \news dataset, after observing cascades for 10 minutes. 
	The \textcolor{blue}{blue} line denotes the overall median ARE shown in Figure~\ref{fig:hybrid-comparison}a.
	Note that the y-axes are not on the same scale for all three graphs.
    }
    \label{fig:performance-popularity}
    \captionmoveup
\end{figure*} 
 
\textbf{Experimental setup.}
Similar to the setup in the previous Sec~\ref{subsec:compare-generative}, we observe cascades for 5 minutes, 10 minutes and 1 hour and fit the \hawkes and \seismic models for each cascade.
The train-test split is different, in order to replicate closer the experimental setup in Martin \etal~\cite{Martin2016}: the data from first half of July (1-15) is used for training and the data from second half of July (16-31) for testing. 
We use the \news historical data from April to June to construct the past user success feature for the feature-driven approach.
We consider only those users who initiated at least 2 cascades in the past as active users.
Non-active users have a past user success of $0$.
We also construct a \hybrid approach, which leverages the features of the \hawkes model -- i.e. $\{ c, \theta, A_1, n^\ast\}$ -- together with the 
features detailed in Sec.~\ref{subsec:data-driven}.
We use a Random Forest regressor for predicting the final volume size after tuning the predictor on the training set with 10 fold cross-validation.  

\textbf{Results.}
Similar to the previous subsection, we report in Table~\ref{tab:feature-driven-comparison} the mean ARE $\pm$ standard deviation and in Fig.~\ref{fig:hybrid-comparison} the box plots of ARE for 10 minutes and 1 hour.
In according with the results of Sec.~\ref{subsec:compare-generative}, \hawkes surpasses \seismic for all considered setups. Note that these results are numerically different from those in Sec~\ref{subsec:compare-generative}, because the underlying test dataset consist only the subset from second half (16-31) of July 2015. 
Surprisingly, even the \featuredriven model outperforms \seismic for all time prefixes, by at least $40\%$. 
In both cases, the differences are statistically significant ($p-val < 0.001$) for 10 minutes and 1 hour.
The \hawkes model exhibits similar mean ARE as \featuredriven and higher standard deviation. 
However, its boxplot in Fig.~\ref{fig:hybrid-comparison} shows lower median and quartiles.
This indicates that \hawkes predict better than \featuredriven for most cascades, but the error distribution of  \hawkes is skewed by outliers. 
\hybrid performs the best, with similar boxplot summarization as \hawkes, but with a $42\%$ better mean APE -- 0.17 to 0.11 for 1 hour. 
The differences between \hybrid and all the other approaches are statistically significant (details in the supplement~\cite{supplemental}). 
This result shows that \hawkes and \featuredriven approaches are complimentary, likely because they 
summarize the cascade information differently, and the errors are uncorrelated. 

\textbf{ARE distribution over cascade popularity.} 
Figure~\ref{fig:performance-popularity} presents the ARE distribution over cascade popularity as boxplots for each popularity decile.
Both \hawkes and \featuredriven predict better for most of the cascades with the middle popularity.
For the extreme ends, performance is affected, because for lower popularity percentiles even a small error is amplified to a large ARE and cascades in higher popularity bins are inherently more difficult to predict.
The prediction performance of \seismic seems consistent across popularity bins, but always worse than \hawkes or \featuredriven, as already showed by the aggregated results in Fig.~\ref{fig:hybrid-comparison}a.
We also tried to explain prediction performance breaking down by \news sources, but we did not see any notable patterns. 

\subsection{Will this cascade double}
\label{subsec:classification-task}

\begin{table}[tbp]
\caption{
	Accuracy $\pm$ standard deviation when predicting whether a cascade will double its size or not.}
\centering
\begin{tabular}{lcc}
\toprule
Approach & 25 tweets & 50 tweets \\ 
\midrule
Random Guess & 0.52 & 0.53 \\ 
\hawkesC & 0.66 $\pm 0.013$ & 0.70 $\pm 0.009$ \\ 
\featuredriven & 0.79 $\pm 0.009$ & 0.81 $\pm 0.011$ \\ 
\hybrid & 0.79 $\pm 0.008$ & 0.82 $\pm 0.013$ \\  
\bottomrule        
\end{tabular} 
\label{tab:classification}
\captionmoveup
\end{table}

The learning problem in this section is to classify whether an observed cascade will double its volume or not.
The experimental setup follows that of Cheng \etal~\cite{CHE14}, in which cascades are observed for a fixed number of retweets, instead of for a fixed extend of time in Sec.~\ref{subsec:compare-generative} and~\ref{subsec:compare-data-driven}.
The proposed \hawkes model outputs directly the total popularity as a cascade size and uses the predictive layer to correct the theoretical estimate.
As a consequence, it cannot be used in a classification setup.
Instead, we construct a classifier \hawkesC, in which a Random Forest Classifier is trained on the same features as the predictive layer of \hawkes and used to output a binary decision.

\textbf{Experimental setup.}
We use the July subset of the \news dataset, filtering to only cascades of length greater than or equal to 25. 
We observe the cascades for two 25 retweets and 50 retweets, and we predict whether they will reach 50 and 100 retweets, respectively. 
We perform a 40\%-60\%  stratified train-test split.
\featuredriven approach uses the features mentioned in Section~\ref{subsec:data-driven}, except for \textit{volume} which is constant for all cascades. 
\hybrid approach combines features from both \hawkes and \featuredriven.
The random train-test split is repeated 10 times and we report mean accuracy and standard deviation.

\textbf{Results.}
Table~\ref{tab:classification} summarizes the classification performances. 
A random guess (majority class) would output an accuracy of $52\%$ and $53\%$ for an observed length of 25 and 50 respectively. 
The generative-based classifier \hawkesC improves substantially over the baseline of random guess, however \featuredriven has the best prediction accuracy.
Interestingly, combining the generative features with the data-driven features did not lead to a significant improvement, likely due to \featuredriven already has stronger results than \hawkesC. 
Admittedly, predicting whether a cascade will double in size is an easier problem than forecasting its final volume.
Overall, the results suggests that the predictions are robust, and the performance are comparable to earlier results reported by Cheng \etal~\cite{CHE14} (0.81 accuracy after seeing the first 50 shares). 



\section{Conclusion}
\label{sec:conclusion}

This paper establishes a common benchmark that allows researchers and practitioners to compare 
 feature-driven and generative approaches on different variants of popularity prediction problems. 
Existing feature-based approaches employ either proprietary insider information, 
or global network information that is impossible to retrieve with the limited access that are openly available. 
In this work, we bridge the problem space by first selecting a small set of features which can be obtained by using only free access services and constructing the \featuredriven method.
We further propose a 
Hawkes self-exciting model, which intuitively aligns with the social factors responsible for diffusion of cascades: social influence of users, social memory and inherent content quality. 
We systematically construct a predictive layer, which helps optimize predict from other cascades.
We perform extensive evaluation on two large datasets: a benchmark dataset constructed by Zhao \etal~\cite{Zhao2015} and a domain-specific dataset curated on news tweets.
We compare the feature-driven approaches and generative approaches in two tasks: i)  predicting the total size of retweeting cascades and ii) predicting whether cascades will double their size.
Both our proposed generative \hawkes method and our \featuredriven method outperform the current state of the art predictor.
Performances are further improved when combining both approaches into \hybrid, which makes us argue that popularity modeling should leverage the best that both worlds have to provide.

We plan to extend this work by including additional factors such as, different network behaviour for particular types of content, user influence distributions that are varying over time, and explicit modeling of the interplay between the unfolding of related cascades.

\vspace{0.2cm}
{ \small
\noindent\textbf{Acknowledgments}
This material is based on research sponsored by the Air Force Research Laboratory, under agreement number FA2386-15-1-4018. 
%
We thank the National Computational Infrastructure (NCI) for providing computational resources, supported by the Australian Government.
}


{ \scriptsize 
\bibliographystyle{abbrv}
\bibliography{paper}  
}
\newpage


\section*{Supplementary material}

# Supporting Information

**Contents**

## 1 Technical details about the optimization procedure

**1.1 Derivation of log-likelihood for the Hawkes process.** We detail the derivation of the log-likelihood in Eq. [5] in the main text:

$$\mathcal{L}(\kappa, \beta, c, \theta) = \log P(\{(m_i, t_i), i = 1, \ldots, n\})$$
$$= \sum_{i=1}^{n} \log(\lambda(t_i)) - \int_{0}^{T} \lambda(\tau) d\tau \quad [13]$$

We start by computing the integral on the right hand side of Eq. [13], denoted by $\Lambda$:

$$\Lambda = \int_{0}^{T} \lambda(\tau) d\tau$$

$$\overset{cf. \, [1]}{=} \int_{0}^{t_1} \overbrace{\sum_{t=0}^{0} \phi_{m_i}(t - t_i)}^{=0} dt +$$
$$\int_{t_1}^{t_2} \phi_{m_0}(t - t_0) dt + \int_{t_1}^{t_2} \phi_{m_1}(t - t_1) dt +$$
$$\int_{t_2}^{t_3} \phi_{m_0}(t - t_0) dt + \int_{t_2}^{t_3} \phi_{m_1}(t - t_1) dt +$$
$$\int_{t_2}^{t_3} \phi_{m_2}(t - t_2) dt +$$
$$\vdots$$
$$\Rightarrow \Lambda = \sum_{i=1}^{n} \int_{t_i}^{T} \phi_{m_i}(t - t_i) dt \quad [14]$$

Introducing Eq. [2] into Eq. [14], we obtain:

$$\Lambda = \sum_{i=1}^{n} \int_{t_i}^{T} \kappa (m_i)^{\beta} \frac{1}{(t - t_i + c)^{1+\theta}} dt$$
$$= \sum_{i=1}^{n} -\kappa (m_i)^{\beta} \frac{(t + c - t_i)^{-\theta}}{\theta} \Big|_{t_i}^{T}$$
$$\Rightarrow \Lambda = \kappa \sum_{i=1}^{n} (m_i)^{\beta} \left[ \frac{1}{\theta c^{\theta}} - \frac{(T + c - t_i)^{-\theta}}{\theta} \right] \quad [15]$$

Hence Equation [13] can be written as,

$$\mathcal{L}(\kappa, \beta, c, \theta) = \overbrace{\sum_{i=2}^{n} \log \kappa}^{1} + \overbrace{\sum_{i=2}^{n} \log \left( \sum_{t_i > t_j} \frac{(m_j)^{\beta}}{(t_i - t_j + c)^{1+\theta}} \right)}^{2} - $$
$$\underbrace{- \kappa \sum_{i=1}^{n} (m_i)^{\beta} \left[ \frac{1}{\theta c^{\theta}} - \frac{(T + c - t_i)^{-\theta}}{\theta} \right]}_{3}$$
$$[16]$$

**1.2 Computation of partial derivatives for optimization.** We present the partial derivatives calculation for Equation 16, used by the Ipopt function optimization procedure for fitting the parameters of the Hawkes self-exciting model to a given cascade.

Differentiating Eq. [16] w.r.t $\kappa$ we obtain:

$$\frac{\partial \mathcal{L}}{\partial \kappa} = \frac{n-1}{\kappa} - \sum_{i=1}^{n} (m_i)^{\beta} \left[ \frac{1}{\theta c^{\theta}} - \frac{1}{(T + c - t_i)^{\theta} \theta} \right] \quad [17]$$

For differentiating $\mathcal{L}(\kappa, \beta, c, \theta)$ w.r.t $\beta$, $c$ and $\theta$, we first consider the $2^{nd}$ second term in the right hand side of Eq. [16]:

$$\sum_{i=2}^{n} \log \left( \sum_{t_i > t_j} \frac{(m_j)^{\beta}}{(t_i - t_j + c)^{1+\theta}} \right) =$$
$$\sum_{i=2}^{n} \log \Big[ (m_{i-1})^{\beta} (t_i - t_{i-1} + c)^{-(1+\theta)} +$$
$$\ldots + (m_1)^{\beta} (t_i - t_1 + c)^{-(1+\theta)} \Big] \quad [18]$$

Using the logarithmic derivation rule and Eq. [18] in Eq [16], we obtain the partial derivative w.r.t $\beta$:

$$\frac{\partial \mathcal{L}}{\partial \beta} = \sum_{i=2}^{n} \frac{\sum_{t_i > t_j} (t_i - t_j + c)^{-(1+\theta)} (m_j)^{\beta} \log(m_j)}{\sum_{t_i > t_j} (t_i - t_j + c)^{-(1+\theta)} (m_j)^{\beta}}$$
$$- \kappa \sum_{i=1}^{n} (m_i)^{\beta} log(m_i) \left[ \frac{1}{\theta c^{\theta}} - \frac{1}{(T + c - t_i)^{\theta} \theta} \right] \quad [19]$$

Similarly, we compute the partial derivatives w.r.t $c$ and $\theta$:

$$\frac{\partial \mathcal{L}}{\partial c} = \sum_{i=2}^{n} \frac{\sum_{t_i > t_j} -(1+\theta) (m_j)^{\beta} (t_i - t_j + c)^{-(2+\theta)}}{\sum_{t_i > t_j} (t_i - t_j + c)^{-(1+\theta)} (m_j)^{\beta}}$$
$$- \kappa \sum_{i=1}^{n} (m_i)^{\beta} \left[ \frac{1}{(T + c - t_i)^{1+\theta}} - \frac{1}{c^{1+\theta}} \right] \quad [20]$$

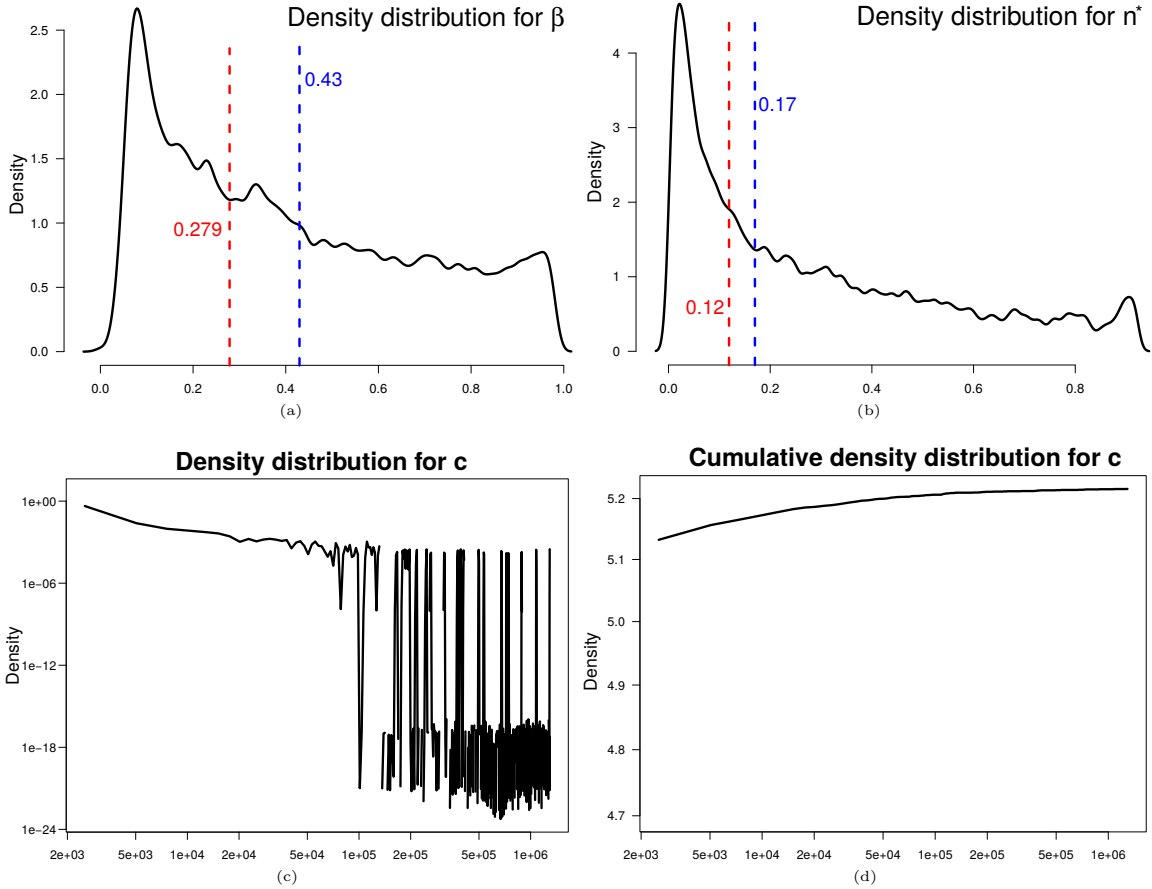

Figure 8: Distribution of parameter $\beta$ (a), $n^*$ (b), $c$ in log-log scale (c) and cumulative distribution of $c$ in log-log (d), on a sample of 17,146 cascades from the News dataset. The dashed vertical lines show fitted parameter values for the two retweet cascades illustrated in the main text Fig. 1: $tw_1$ in red and $tw_2$ in blue.

$$\frac{\partial \mathcal{L}}{\partial \theta} = \sum_{i=2}^{n} \frac{\sum_{t_i > t_j} -\log(t_i - t_j + c)(t_i - t_j + c)^{-(1+\theta)}(m_j)^\beta}{\sum_{t_i > t_j}(t_i - t_j + c)^{-(1+\theta)}(m_j)^\beta}$$
$$- \kappa \sum_{i=1}^{n} \frac{(m_i)^\beta}{\theta^2} \left[ (\theta \log(T + c - t_i) + 1)(T + c - t_i)^{-\theta} - (\theta \log(c) + 1) c^{-\theta} \right] \quad [21]$$

## 2  Additional density distribution graphics

Figure 8 shows the density distribution of parameters $\beta$ and $c$, along with branching factor $n^*$. The density of $\beta$ shows two local maxima: a maximum of density around 0.079 and a smaller peak around $\beta \sim \alpha - 1$ resulting from the optimizer terminating at the boundary of the non-linear constraint $n^* = 1$.

Parameter $n^*$ represents the expected number of events spawned by a single event in the Hawkes process. Its distribution shows a maximum around $n = 0.022$ and a smaller local maximum around 1, resulted from the same process involving the non-linear constraint $n^* = 1$. The low value of $n^*$ point towards a fast decaying self-exciting process, resulting in small cascades with short life spans.

Parameter $c$ is a cutoff parameter, its purpose is to keep the kernel function $\phi_m(\tau)$ bounded when $\tau \sim 0$. It controls the *reactivity* of the initial diffusion: small values lead to the first retweets being posted early after the initial tweet; large values introduce long waiting times between the first tweet and subsequent retweets. Fig. 8(c) shows its long-tail density distribution in log-log scale and Fig 8(d) shows its cumulative density distribution. The large majority of cascades are very reactive, with a median value of $c = 111$ seconds.

Table 5: Mean and Median values of log-likelihood of unseen data on a sub sample of News dataset with 100 cascades for Seismic and unconstrained version of Hawkes.

| Approach | Mean | Median |
|---|---|---|
| Seismic | -6.620 | -4.3964 |
| Hawkes | -7.050 | -6.620 |

## 3  Goodness of Fit

As we have a generative model that predicts final size of a cascade better than the current state of the art [1], we set

out the task to evaluate how good our model fits the unseen data. In order to achieve this we first fit a model to all data seen till some time $t_N$ for a cascade and then estimate the log-likelihood of the unseen data till it's end time $t_M$.

We can get the log-likelihood of the unseen data by subtracting the log-likelihood of the seen cascade from the log-likelihood of the whole cascade after fitting the parameters on the seen part, it is defined as follows,

$$\mathcal{L}(\{(m_i, t_i), i = N+1, \ldots, M\}) = \mathcal{L}(\{(m_i, t_i), i = 1, \ldots, M\}) \\ - \mathcal{L}(\{(m_i, t_i), i = 1, \ldots, N\}) \quad [22]$$

For HAWKES we can easily calculate the value in Equation 22 by using Equation 16 for both seen and full cascade. But for SEISMIC the evaluation is not possible, as they do not derive the log-likelihood of their model explicitly. For SEISMIC the derivation of log-likelihood of unseen data can be written as follows,

$$\mathcal{L}(\{(m_i, t_i), i = N+1, \ldots, M\}) = \overbrace{\sum_{i=N+1}^{M} \log(\lambda(t_i))}^{A} \\ - \underbrace{\int_{t_N}^{t_M} \lambda(\tau) \mathrm{d}\tau}_{B} \quad [23]$$

Now we can calculate the part A of Equation 23 by using Equation 1 in [1] and assuming that infectiousness once estimated at $t_N$ remains constant, as follows,

$$A = \sum_{i=N+1}^{M} \log(P_{t_N}) + \sum_{i=N+1}^{M} \log\left(\sum_{j=1}^{i} n_j \phi(t_i - t_j)\right) \quad [24]$$

Now for Part B in Equation 23 by using Equation 1 in [1], we have,

$$B = \int_{t_N}^{t_M} P_{t_N} \sum_{t_i \leqslant \tau} (\phi(\tau - t_i) n_i \mathrm{d}\tau) \quad [25]$$

Now using the trick used in Equation 14 and keeping in mind the fact that $t_i$ has two parts, i$\in [1, N]$ and i$\in [N+1, M]$, we can write Equation 25 as,

$$B = P_{t_N} \left[\sum_{i=1}^{N} n_i \int_{t_N}^{t_M} \phi(\tau - t_i) \mathrm{d}\tau \\ + \sum_{i=N+1}^{M} n_i \int_{t_i}^{t_M} \phi(\tau - t_i) \mathrm{d}\tau\right] \quad [26]$$

Using Equation 24 and Equation 26 we can calculate the desired log-likelihood of the unseen data for SEISMIC.

To present results for goodness of fit we use unconstrained optimization of parameters for HAWKES and compare it against SEISMIC on a sub sample of NEWS dataset with 100 cascades. The results are laid down in Table 5. As seen from the table, SEISMIC fits slightly better to the unseen data HAWKES, even though the predictive power of HAWKES is better as demonstrated in main text.

## 4 More Results

In this section we present the analysis on statistical significance of our results presented in Section 6.2 of main text.

Results are presented in Table 6 for SEISMIC, HAWKES, FEATURE-DRIVEN and HYBRID models for the NEWS dataset, we use a strict cut-off p-value, 0.01, as our sample size is in order of thousands. As seen from Table 6 HYBRID model shows statistical significant improvement over all other approaches almost every time. The results for HAWKES and FEATURE-DRIVEN are in general better than SEISMIC, but statistically insignificant among each other.

Table 6: Statistical testing results for SEISMIC, HAWKES, FEATURE-DRIVEN and HYBRID models for the NEWS dataset, indicating the algorithm with statistical significant better results. Win is decided when an algorithm has majority of lower error values than the competing algorithm and p-value confirms the results as statistically significant, else we mark it a draw in case of no statistical difference.

| Approach | 5 minutes | 10 minutes | 1 hour |
|---|---|---|---|
| SEISMIC-FEAT.-DRIVEN | Draw | FEAT.-DRIVEN | FEAT.-DRIVEN |
| SEISMIC-HAWKES | Draw | HAWKES | HAWKES |
| SEISMIC-HYBRID | Draw | HYBRID | HYBRID |
| FEAT.-DRIVEN-HAWKES | Draw | Draw | Draw |
| FEAT.-DRIVEN-HYBRID | HYBRID | HYBRID | HYBRID |
| HAWKES-HYBRID | HYBRID | HYBRID | HYBRID |


\end{document}